\documentclass[12pt,preprint]{aastex}
\usepackage{soul}
\usepackage{ulem}

\newcommand{\Msun}{\hbox{M$_{\odot}$}}

\newcommand{\ha}{H$\alpha$}
\newcommand{\HST}{{\sl HST}}

\newcommand{\rosat}{{\sl ROSAT}}

\newcommand{\galex}{{\sl GALEX}}

\slugcomment{Accepted to ApJ}

\shorttitle{Searching for Young M Dwarfs with \galex}
\shortauthors{Shkolnik et al.}

\begin{document}


\title{Searching for Young M Dwarfs with \galex\ \altaffilmark{1}\\}


\author{Evgenya~L.~Shkolnik\altaffilmark{2}}
\affil{Department of Terrestrial Magnetism, Carnegie Institution of Washington, 5241 Broad Branch Road, NW, Washington, DC 20015}
\email{shkolnik@dtm.ciw.edu}

\author{Michael C. Liu}
\affil{Institute for Astronomy, University of Hawaii at Manoa\\ 2680 Woodlawn Drive, Honolulu, HI 96822}
\email{mliu@ifa.hawaii.edu}

\author{I. Neill Reid}
\affil{Space Telescope Science Institute, Baltimore, MD 21218}
\email{inr@stsci.edu}

\author{Trent Dupuy}
\affil{Institute for Astronomy, University of Hawaii at Manoa\\ 2680 Woodlawn Drive, Honolulu, HI 96822}
\email{tdupuy@ifa.hawaii.edu}

\and

\author{Alycia J. Weinberger}
\affil{Department of Terrestrial Magnetism, Carnegie Institution of Washington, 5241 Broad Branch Road, NW, Washington, DC 20015}
\email{alycia@dtm.ciw.edu}

\altaffiltext{1}{This paper is based on data gathered with the 6.5 m Magellan Telescopes
located at Las Campanas Observatory, Chile, the Keck II telescope, and the \galex, 2MASS and HST/GSC v2.3 photometric catalogs. GALEX is operated for NASA by the California Institute of Technology under NASA contract NAS5-98034.}
\altaffiltext{2}{Carnegie Fellow}

\begin{abstract}

	\noindent The census of young moving groups in the solar neighborhood is
	significantly incomplete in the low-mass regime. We have developed a new
	selection process to find these missing members based on the \galex\
	All-Sky Imaging Survey (AIS). For stars with spectral types $\gtrsim$K5 ($R-J \gtrsim 1.5$)
	and younger than $\approx$300~Myr, we show that near-UV (NUV) and far-UV
	(FUV) emission is greatly enhanced above the quiescent photosphere,
	analogous to the enhanced X-ray emission of young low-mass stars seen by
	\rosat\ but detectable to much larger distances with \galex.
	By combining \galex\ data with optical (HST Guide Star Catalog) and
	near-IR (2MASS) photometry, we identified an initial sample of 34 young
	M dwarf candidates in a 1000 sq.~deg.~region around the $\approx$10-Myr TW Hydra
	Association (TWA). Low-resolution spectroscopy of 30 of these found 16 which had
	\ha\ in emission, which were then followed-up at high resolution to search
	for spectroscopic evidence of youth and to measure their radial
	velocities.
	Four objects have low surface gravities, photometric distances and space
	motions consistent with TWA, but the non-detection of Li indicates they
	may be too old to belong to this moving group.
	One object (M3.5, 93$\pm$19~pc) appears to be the first known accreting
	low-mass member of the $\approx$15~Myr Lower Centaurus Crux OB
	association.
	Two objects exhibit all the characteristics of the known TWA members,
	and thus we designate them as TWA 31 (M4.2, 110$\pm$11~pc) and TWA 32
	(M6.3, 53$\pm$5 pc).  TWA 31 shows extremely broad (447 km~s$^{-1}$)
	H$\alpha$ emission, making it the sixth member of TWA found to have
	ongoing accretion. TWA 32 is resolved into a 0.6\arcsec\ binary in Keck
	laser guide star adaptive optics imaging.
	Our search should be sensitive down to spectral types of at least M4--M5 in TWA and thus the small numbers of new member is puzzling.  This might indicate TWA has an atypical mass function or that the presence of lithium absorption may be too restrictive a criteria for selecting young low-mass stars.

\end{abstract}


\section{Introduction}\label{intro}

Observational studies of planet formation have been energized
by the discovery of young ($<$100 Myr) solar-type stars close to
Earth (e.g.~\citealt{jeff95,webb99}), identified from multiple indicators of youth
including chromospheric activity and strong X-ray emission.  The
combination of distances, proper motions, and radial velocities (RVs) has
allowed many of these stars to be kinematically linked to coeval
moving groups (e.g., \citealt{zuck04,torr08}).  These young moving groups (YMGs)
are several times closer to Earth than the traditional
well-studied star-forming regions such as Taurus and Orion
($\sim$150--500 pc).  More importantly, these groups have ages of
$\sim$10--100 Myr, a time period in stellar evolution that has  been largely underrepresented
in previous studies.  This is expected to be a key epoch for
understanding planet formation, coinciding with the end of giant planet
formation and the active phase of terrestrial planet formation (e.g.~\citealt{mand07,ida08}).

\begin{figure}
\epsscale{.60}
\plotone{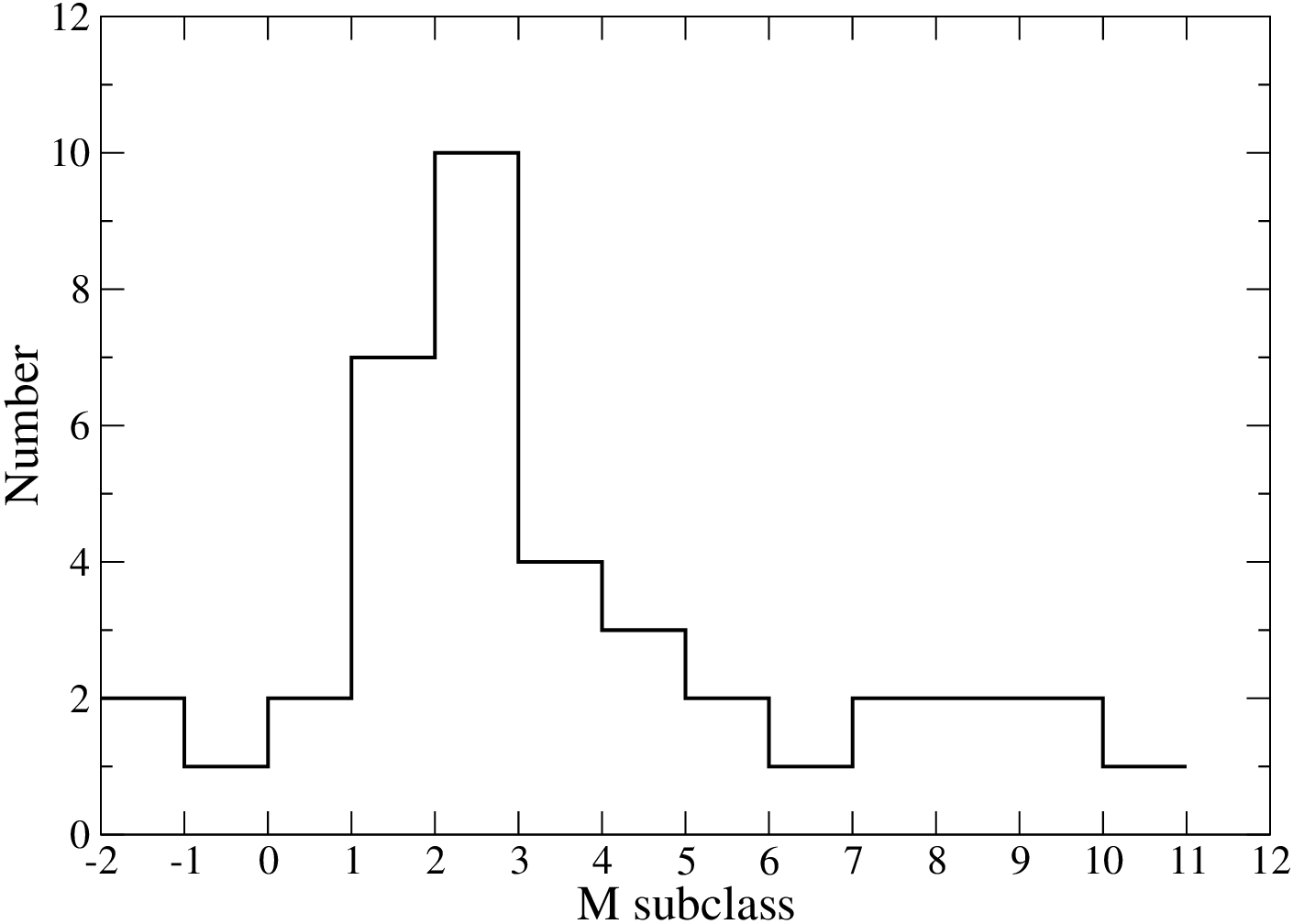}
\caption{Spectral type histogram of known TWA members including TWA 31 and 32.\newline \label{histogram}}
\end{figure}

\begin{figure}
\epsscale{.60}
\plotone{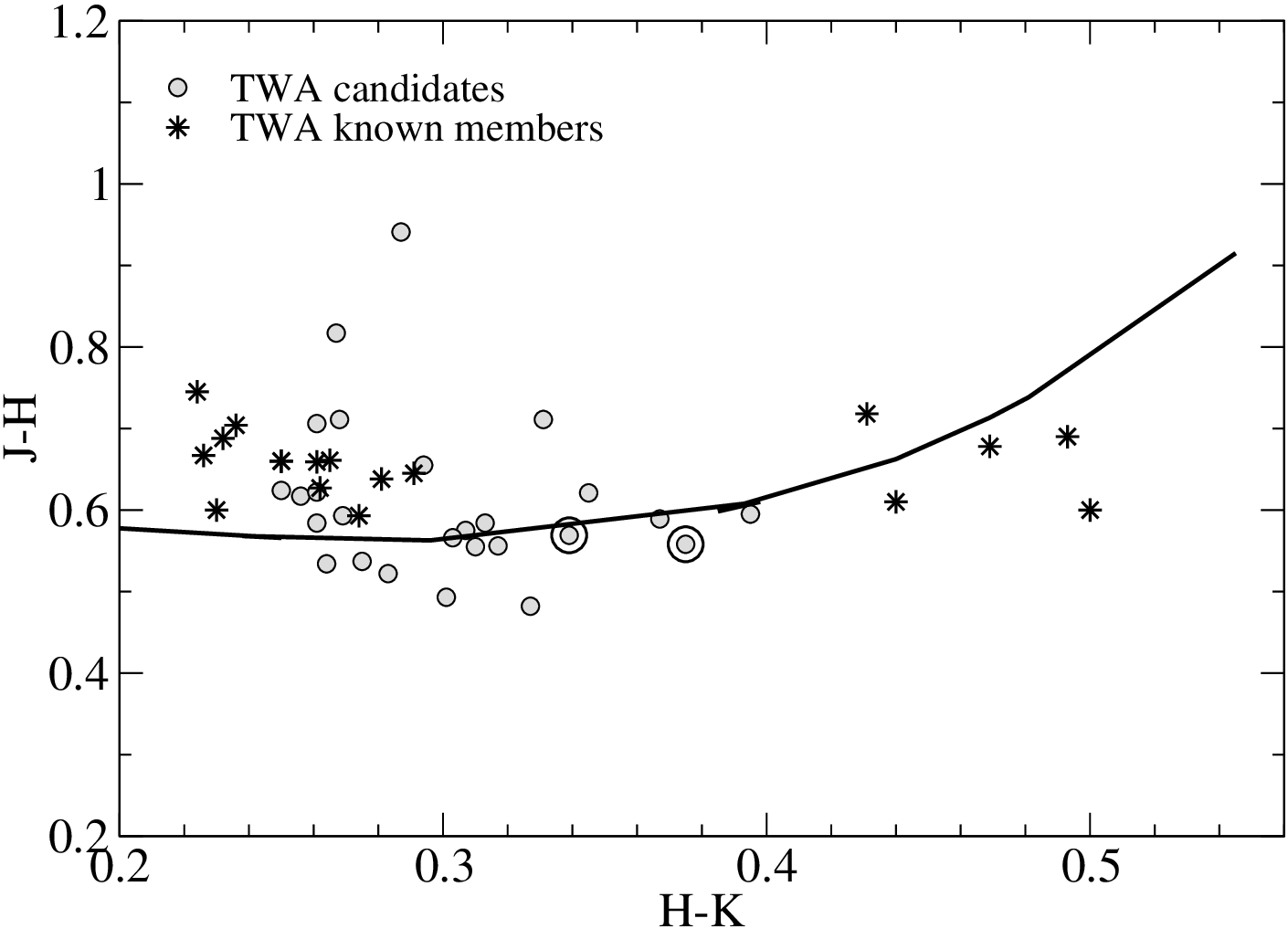}
\caption{A 2MASS color-color plot of our TWA candidates plus known TWA members. The two stars circled are TWA 31 and TWA 32, and are discussed in detail in Section~\ref{newmembers}. Neither shows significant K-band excess using the M dwarf color sequence of \cite{cush05}.\newline \newline\newline\label{colour_colour}}
\end{figure}

One of the best studied of these YMGs is the TW Hydrae Association (TWA; \citealt{kast97}).  Its unique combination of youth
($\sim$10 Myr; \citealt{barr06,ment08}), proximity ($\sim$30--100 pc) and (relative) compactness on the sky, makes it is a particularly promising observational test bed for theories describing disk
evolution and planet formation.
TWA is named for the actively accreting K7 star with a circumstellar disk (\citealt{heni76,ruci83,reza89}). The discovery of such a young star far from any star-forming region prompted several searches for additional T Tauri stars in the same part of the sky. These searches used infrared excesses with IRAS (\citealt{greg92}), \rosat\ X-ray activity (\citealt{ster99,webb99}), and common space motion (\citealt{song03,scho05}), plus near-IR photometric searches for brown dwarfs (\citealt{gizi02,loop07}). The current census of accepted TWA members is up to 25 objects (plus 2 discovered by us), including either late-K or early-M stars and a few brown dwarfs, making TWA seem unusual in its initial mass function compared to other star forming regions (e.g.~\citealt{torr08,sles08}).  In particular, there appears to be a dearth of M3--M7 stars (Figures~\ref{histogram} and \ref{colour_colour}; \citealt{mama05} and references therein).  It is possible that this large
incompleteness arises from the fact that most young star searches have mostly relied on bright optical catalogs
(e.g., Hipparcos) as a starting point, which are deficient in M dwarfs, or perhaps TWA may have an unexpected mass function.

Over the last few years, we have been undertaking a systematic effort to complete the young low-mass census through multi-catalog selection and followup spectroscopy. Our initial work
focused on the immediate solar neighborhood ($<$25 pc), identifying the youngest M dwarfs from the known (Gliese \& NStars) catalogs. Our spectroscopy had a very productive
($\approx$80\%) confirmation rate, identifying 144 new young ($\lesssim$ 300 Myr) M dwarfs using X-ray selection (\rosat\ Bright Source Catalog; \citealt{voge99}) as a first cut (\citealt{shko09b}, hereafter SLR09; \citealt{shko08b, shko10}). 
By scrutinizing the \galex\ NUV and FUV properties of this 25-pc sample, we have constructed
efficient selection criteria to expand the search for young M dwarfs to 100 pc thereby making the M dwarfs of many of the YMGs accessible. In this paper, we apply our new methodology to TWA to find its `missing' M stars.

\section{Selection of Young Stars using the \galex\ All-sky Imaging Survey}

Stellar activity is a
well-established indicator of youth (e.g., \citealt{prei05}), and the only all-sky surveys suited for finding active,
young stars have been the \rosat\ X-ray catalogs (e.g.~Voges et al.~1999). However, since the luminosities of M dwarfs are
$\sim$10--300$\times$ lower than solar-type stars, \rosat\ is generally limited to the nearest, earliest-type M dwarfs.

The NASA Galaxy Evolution Explorer (\galex; \citealt{mart05}) provides a new resource that enables a major expansion of the
young low-mass census, far beyond previous data sets. The \galex\ satellite was launched on April 28, 2003 and has imaged most of
the sky simultaneously in two bands: near-UV (NUV) 1750--2750 \AA\ and far-UV (FUV) 1350--1750 \AA, with angular
resolutions of 5\arcsec and 6.5$\arcsec$ across a 1.25$^{\circ}$ field of view. The full description of the instrumental
performance is presented by \cite{morr05}. The \galex\ mission is producing an All-sky Imaging Survey (AIS) which is archived at
the Multi-mission Archive at the Space Telescope Science Institute (MAST).\footnote{One can query the AIS through either
CasJobs (http://mastweb.stsci.edu/gcasjobs/) or a web tool called GalexView (http://galex.stsci.edu/galexview/).} The NUV/FUV
fluxes and magnitudes are produced by the standard \galex\ Data
Analysis Pipeline (ver.~4.0) operated at the Caltech Science Operations Center \citep{morr05}. The data presented in this
paper made use of the fourth data release of the AIS (GR4), covering 2/3 of the sky.\footnote{See details at http://www.galex.caltech.edu/researcher/techdoc-ch2.html.} 

For M dwarfs, the flux in the \galex\ bandpasses is made up of an abundance of
emission lines. Stellar flare activity on M dwarfs observed by \cite{robi05} and \cite{wels07} has
shown that \galex's FUV flux is mainly due to transition region C IV ($\sim$50\%), while weaker
emission lines and continuum provide the remaining 50\%. In the NUV band, flux is primarily due to
continuum emission plus Mg II (10\%), Fe II (17\%), Al II and C III (14\%) lines. In addition,
coronal lines provide 2\% of the FUV and 10\% of the NUV emission.~(See also
\citealt{paga09}.) This makes \galex\ ideal for finding active low-mass stars (e.g.~\citealt{find10}). 

We first tested \galex's sensitivity for finding young M dwarfs by
correlating the X-ray-selected sample of M dwarfs within 25 pc described in SLR09
with the \galex/AIS archive.
Sixty-seven percent were detected in the NUV band, corresponding to the
67\% sky coverage of the AIS in the GR4 data release, i.e.~all those observed by \galex\ were detected.  
This indicates that \galex\ is {\it at least} as sensitive as
\rosat\ in identifying young M dwarfs. The GR4 release also
detected 17 of the 25 (68\%) known M dwarfs in TWA
(not double-counting visual binaries (VBs) not resolved by \galex), with 7 of the 8
remaining not yet observed. The only TWA members observed by \galex\ but not confidently detected were TWA~30 A and B \citep{loop10a, loop10b}.\footnote{\galex\ does have a weak (2.4$\sigma$) detection in the NUV bandpass 26$\arcsec$ away from TWA~30 A's 2MASS coordinates, with no corresponding FUV detection. The non-detection of TWA 30 A may be related to the unusually low \ha\ EW and spectral variability observed for this target. \cite{loop10a} speculate this is due to an accretion disk viewed nearly edge-on with the stellar rotation axis inclined to the disk. Similarly, there is an 8$\sigma$ detection 27$\arcsec$ away in a different direction from TWA 30 B \citep{loop10b}.}

\begin{figure}
\epsscale{.60}
\plotone{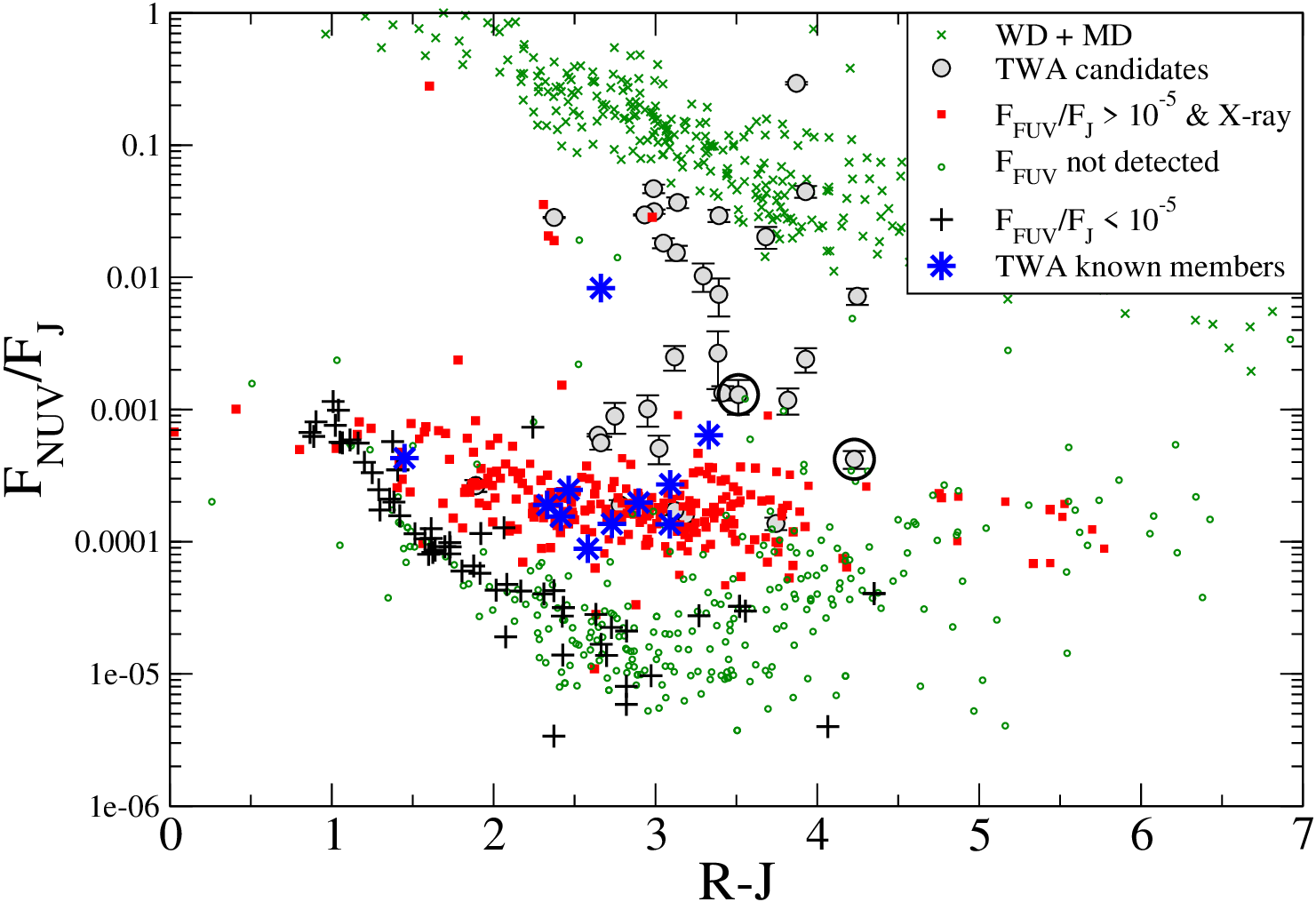}
\caption{Fractional near-UV flux plotted
    against $R-J$ for the \galex-detected M dwarfs, including those from the NStars 25-pc list (\citealt{reid07b}). The
    red squares represent X-ray-bright stars within 50 pc
    ($\lesssim$ 300 Myr; \citealt{shko09b, riaz06}). 
     The TWA candidates targeted for spectroscopic followup have both high NUV and FUV fractional luminosities. However those with $F_{NUV}/F_J \gtrsim 0.01$ with no \ha\ emission are most likely M dwarf + white dwarf pairs (green crosses; \citealt{silv07}). The two new TWA members reported here are identified with large circles. 
\label{RJ_NUV}}
\end{figure}

We then carried out a search for UV detections with the NStars
25-pc census ($\approx$1500 M dwarfs; Reid et al.~2007).  Those with NUV detections are plotted
in Figure~\ref{RJ_NUV} as a function of $R-J$ color as a proxy for effective
temperature.  
From the results of this \galex\ 25-pc analysis, we found that NUV
  data yield many candidates, but applying FUV criteria provide
  an excellent means to distinguish between the
  (never-before-delineated) quiescent emission of old stars
  ($F_{FUV}/F_J$ $<$ 10$^{-5}$), 
  the faint sources
  (FUV not detected), 
  and the truly young targets with high levels of NUV and FUV
  emission, where $F_{NUV}$, $F_{FUV}$ and $F_{J}$ are the stellar fluxes in the respective bandpasses. 
We find that for $R-J\lesssim$4 (SpT
$\lesssim$ M4), we can detect young M dwarfs at least out to several hundred parsecs.  And for young stars later than
$R-J$$\approx$6 (SpT $\sim$ M7), we can probe out to distances of $\approx$75 -- 100 pc and further for very active stars.  Thus \galex\
offers an enormous advantage for detecting young M dwarfs compared to
\rosat.

In principle, flaring M dwarfs could contaminate our sample since UV flares would be mistaken for strong steady-state UV
excesses.  However, we expect the contribution of these objects to be
less than 3\%, based on the fraction of short-term flaring M dwarfs
from \galex\ results thus far (Welsh et al.~2006).  

To summarize, the \galex/AIS is substantially more sensitive than
  previous \rosat\ catalogs, meaning that young M dwarfs can be identified at much greater distances and to
  later spectral types.  In this study, we use the \galex\ NUV and FUV data combined with existing optical and NIR photometric catalogs to identify TWA candidate members within 100 pc of the Sun.\footnote{While this manuscript was under review, a preprint by \cite{rodr10} presented a NUV+NIR search for members of TWA and the Scorpius-Centaurus, including the star we refer to as TWA 32. However, their RV and UVW of this star are inconsistent with our measurements. They exclude it as a member of any known YMG.}

\section{Sample Selection of TWA Candidates}\label{selection}

\begin{figure}
\epsscale{0.6}
\plotone{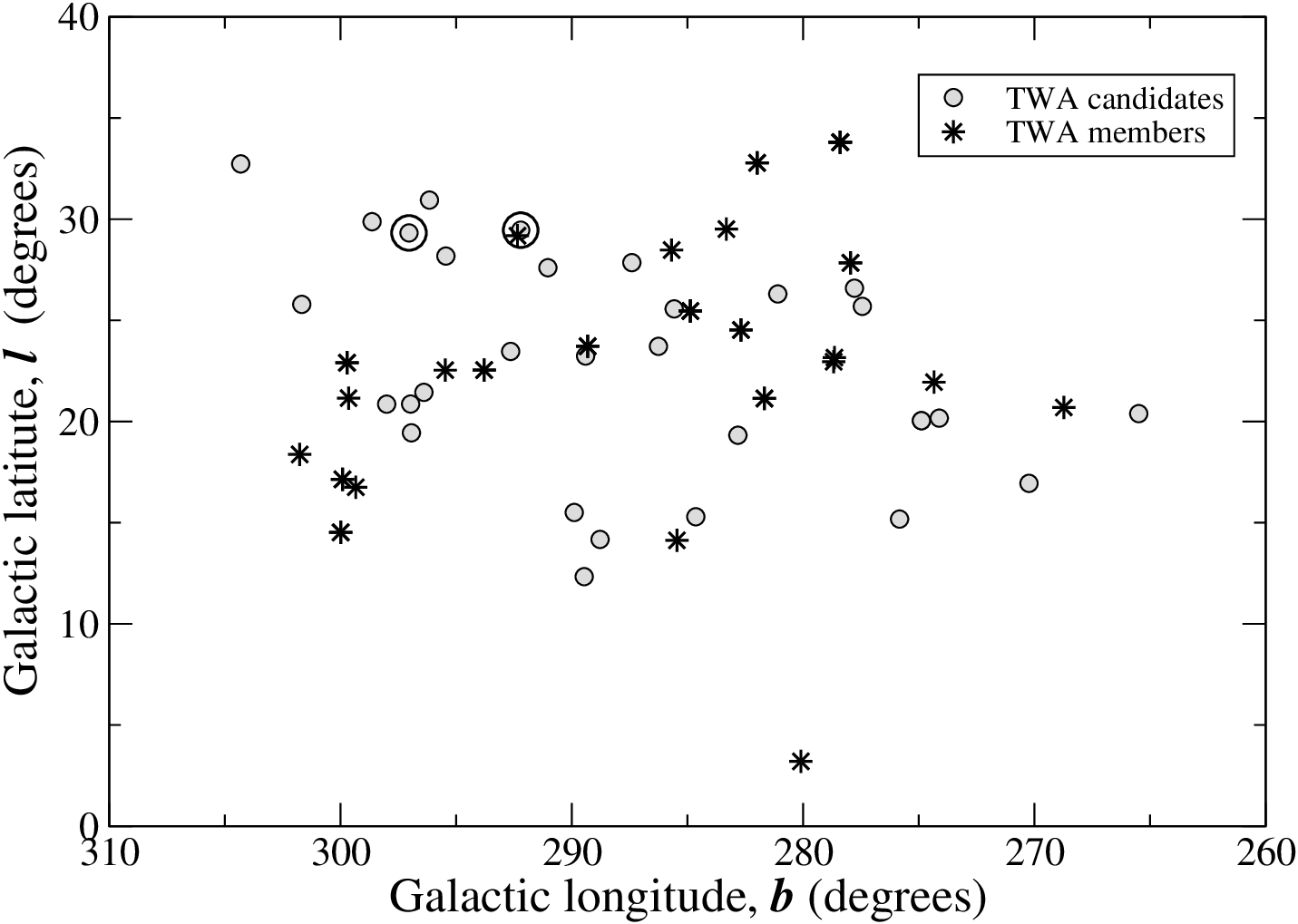}
\caption{Galactic latitude and longitude of known TWA members and our candidates.\label{glon_glat}}
\end{figure}

Late-K and M~dwarfs have distinctive photometric properties, and we can use these to identify stars within 100~pc. We combined optical photometry from the \HST\ Guide Star Catalogue
(GSC 2.3; \citealt{lask08}) with $JHK$ photometry from the 2MASS Point Source Catalogue (\citealt{cutr03}) to identify candidate late-type dwarfs.
We first queried the 2MASS catalog for all sources with $H-K$ $>$ 0.25 (Figure~\ref{colour_colour}), aiming to include everything with SpT later than M2 within a 1000 sq.~deg.~region around the known TWA members bounded by these position limits:  RA={10 to 13} hrs, DEC={--30 to --60} degrees, $b >$ 10 degrees (Figure~\ref{glon_glat}). This yielded 261,547 objects. Cross-matching these against the GSC 2.3 catalog returned 183,361 targets. A 10\arcsec\ cross-matching with the \galex/AIS returned 1968 targets with $>$3$\sigma$ detections in both the NUV and FUV, as well as eliminated any background early-M giants.

\begin{figure}
\epsscale{.60}
\plotone{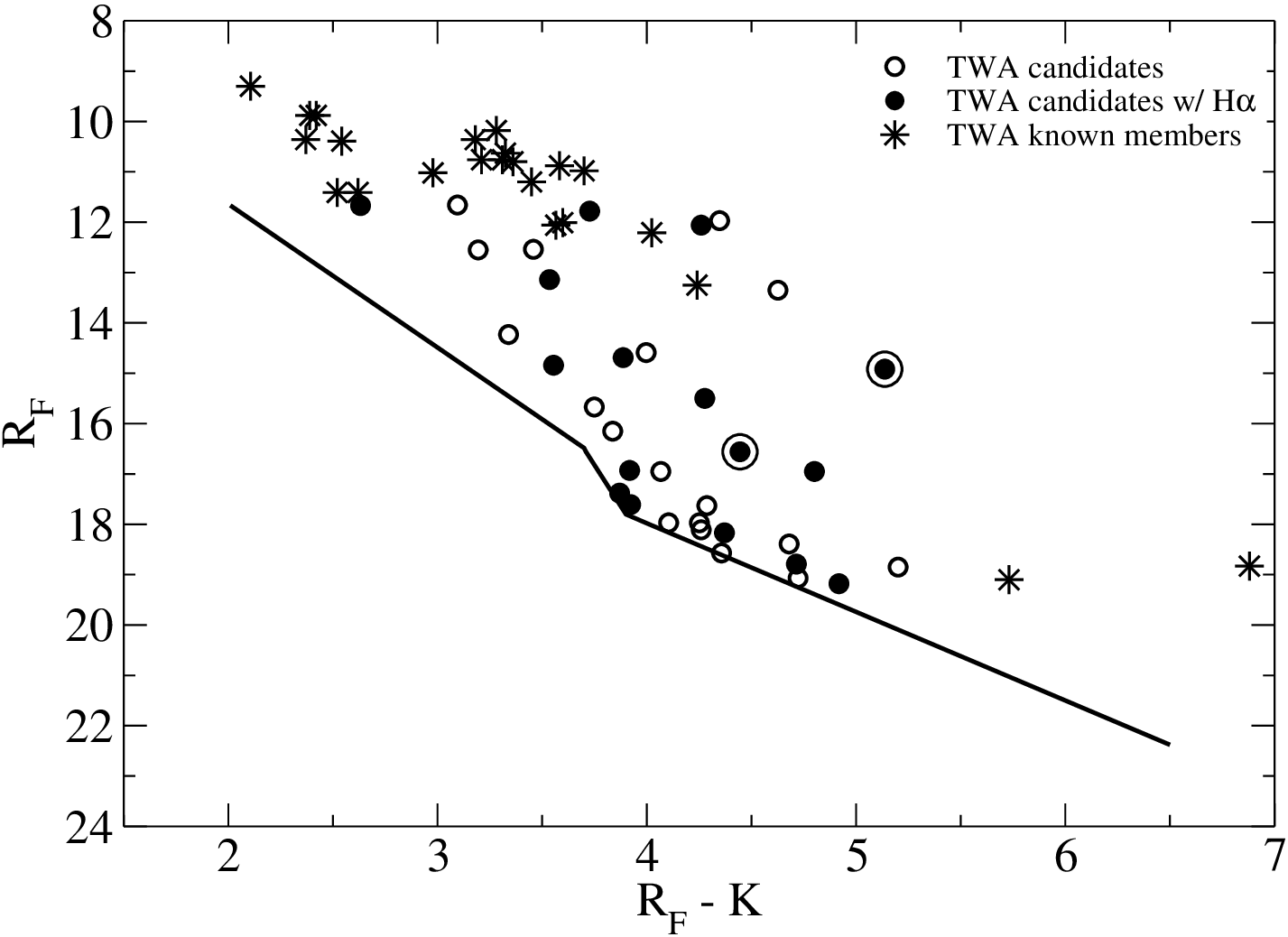}
\caption{A color-magnitude diagram showing the TWA candidates for which we collected low-resolution data.  The lines are the applied color and magnitude cuts detailed in Section~\ref{selection} which set a photometric distance limit of 100 pc (assuming a sample of main-sequence stars).\newline \label{colour_mag}}
\end{figure}

The distance limit was set using an optical/near-infrared color-magnitude diagram, $R_F$ vs.~$R_F$--$K$ (Figure~\ref{colour_mag}, \citealt{reid07b}), where $R_F$ is GSC 2.3's photometric optical band with a wavelength range from 6000 to 7500 \AA, very similar to the standard $R$ passband. This method has proven extremely effective at identifying cool dwarfs within 25 pc, and
it was a simple matter to extend it to larger distances. For young pre-main-sequence (PMS) stars which are over-luminous compared to dwarfs, the effective distance limit becomes 120 pc for the early M's 
based on PMS models by \cite{bara98}. 
Our resulting photometric cuts were:\\

~~~~~~~~~~~~~~~~~~~~~~$R_F$ $<$ 2.85($R_F$--$K$)+5.94;~~ 2 $<$ ($R_F$--$K$) $\leq$ 3.7

~~~~~~~~~~~~~~~~~~~~~~$R_F$ $<$ 6.5($R_F$--$K$)--7.55;~~ 3.7 $<$ ($R_F$--$K$) $\leq$ 3.9

~~~~~~~~~~~~~~~~~~~~~~$R_F$ $<$ 1.76($R_F$--$K$)+10.94;~~ 3.9 $<$ ($R_F$--$K$) $\leq$ 6.5\\

Fifty-five targets remained after applying these color-cuts.  The TWA candidates chosen have both high NUV and FUV fractional luminosities ($F_{FUV}/F_J$ $>$ 10$^{-5}$ and $F_{NUV}/F_J$ $>$ 10$^{-4}$), comparable to or greater than strong X-ray emitting M dwarfs (SLR09) and known TWA members. This generated a target list of 38 stars (including 4 previously known TWA members, TWA 3A, 3B, 10 and 12)\footnote{The known TWA members not listed here were outside of our search criteria, i.e.~had $H-K <$ 0.25, had declinations above --30 degrees, or, as is the case for the brown dwarf members, were not detected in the FUV bandpass at the 3$\sigma$ level.} for spectroscopic followup (Figures~\ref{RJ_NUV} and \ref{colour_mag}).

\section{Spectroscopic Confirmation}\label{spectra}

We then subjected our compiled list of \galex-selected TWA candidates through a two-step process of ground-based spectroscopy to
assess their potential membership in TWA. We first acquired low-resolution spectra to search for the activity-sensitive emission
line \ha\ always present in young low-mass stars ($\lesssim$400 Myr; \citealt{west08}). Thirty of the 34 candidates were observed
with the Magellan Echellette Spectrograph (MagE; \citealt{mars08}) mounted on the 6.5-m Clay telescope on UT 2009 February 10 (Table~\ref{table_phot}). MagE
is a moderate resolution, cross-dispersed echellette, covering the optical wavelengths from 3100 -- 10000 \AA. We used the
0.$\arcsec$7 slit which produced a resolution of 4100. Wavelength calibration was performed with the ThAr exposures taken before and
after each stellar target.

Of these 30, 16 exhibited \ha\ in emission (EW $<$--1 \AA), for which we acquired high-resolution spectra using the Magellan Inamori Kyocera Echelle (MIKE)
spectrograph also at the Clay telescope over several nights: UT 2009 April 15, June 6 - 8, December 31 and 2010 January 1. We also
observed 14 of the known TWA M dwarfs\footnote{Our spectrum of TWA 23 revealed it to be a double-lined spectroscopic binary (SB2).} (Tables~\ref{twa_members} and \ref{table_twamembers_kinematics}). We used the 0.5$\arcsec$ slit which produces a spectral resolution
of $\approx$35000 across the 4900 -- 10000 \AA\/ range of the red chip.
These data were reduced using the facility pipeline \citep{kels03}. Each stellar exposure was bias-subtracted and flat-fielded for pixel-to-pixel sensitivity variations. After optimal extraction, the 1-D
spectra were wavelength calibrated with a ThAr arc. To correct for
instrumental drifts, we used the telluric molecular oxygen A band (from
7620 -- 7660 \AA) which aligns the MIKE spectra to 40 m~s$^{-1}$, after
which we corrected for the heliocentric motion of the Earth. The final spectra are of
moderate S/N  ($\approx$ 25 per pixel at 8000 \AA).

The MIKE data provide RV measurements to better than 1 km~s$^{-1}$ in almost all cases.  
We measured RVs of our 16 candidates and 14 known TWA members by cross-correlating their spectra with those of two RV standards taken on the same night, namely GJ 179 (SpT = M4; \citealt{marc87}) and/or GJ 908 (SpT = M1; \citealt{nide02}). We cross-correlated each of 9 spectral orders 
between 7000 and 9000~\AA\/ (excluding those orders with strong telluric absorption)
where M dwarfs emit most of their optical light using IRAF's\footnote{IRAF (Image
  Reduction and Analysis Facility) is distributed by the National Optical
  Astronomy Observatories, which is operated by the Association of
  Universities for Research in Astronomy, Inc.~(AURA) under cooperative
  agreement with the National Science Foundation.} {\it fxcor} routine
(\citealt{fitz93}).   We measured the RVs from the gaussian peak fitted to the
cross-correlation function of each order and adopted the average RV of all
orders. The target RVs and their standard deviations are listed in
Table~\ref{table_spec}.

The high-resolution spectra also allowed us to identify multi-lined binaries whose excess UV emission may be due to the tidal locking of a close-in binary system rather than youth.
We found 2 (possibly 3) of the 16 to be SB2s consistent with the 16\% low-mass spectroscopic binary fraction of our \rosat\ study (SLR09, \citealt{shko10}). With single epoch observations we are unable to exclude SB1s from the remaining sample, however, given the results of SLR09, we do not expect any to be found with orbital periods less than 1--2 years.

\section{Spectral Types}\label{spt}

The spectra of M dwarfs are dominated by the strong TiO molecular bands, which are particularly diagnostic of the star's temperature.
To estimate spectral types (SpT) we used the TiO-7140 index defined by \citet{wilk05} as the ratio of the mean flux in two 50-\AA\/ bands: the `continuum' band centered on 7035 \AA\/ and the TiO band on 7140 \AA. 
We calibrated the TiO-7140 index for our data sets using 126 M dwarfs from the NStars 25-pc sample (\citealt{reid07b})  we observed with MIKE that have published spectral types plus several RV standards and known members of TWA (\citealt{reid95,riaz06}, \citealt{weng07} and references therein).  
The relationship we used to convert the TiO-7140 index to SpT for M0 -- M6 stars is: 

\begin{center}
SpT=(TiO$_{7140}-1.1158)/0.1802$, {\it rms}= 0.5 subclasses~~~~~~~~(1)
\end{center}

\noindent Here, M0 corresponds to SpT=0, M1 $\rightarrow$ 1, M2 $\rightarrow$ 2, etc. The linear fit to the calibration is consistent to better than 0.2 subclasses with that used for the Keck/HIRES and CFHT/ESPaDoNS data presented in SLR09. 
We determined the errors of our measurements for the TiO index (and subsequent indices discussed below) by taking the mean standard deviations measured for 5 RV standards stars observed with the same setup on multiple nights. We calculated an average error of 0.012 in the TiO index measurement, which translates to only 0.07 subclasses in SpT.  
Although this uncertainty is small, the calibration is based on a sample with SpTs binned to half a subclass, imposing a 0.5 subclass uncertainty in the calculated SpTs listed in Table~\ref{table_phot}.

\section{Identifying the Young Stars} \label{youngstars}

\begin{figure}
\epsscale{.60}
\plotone{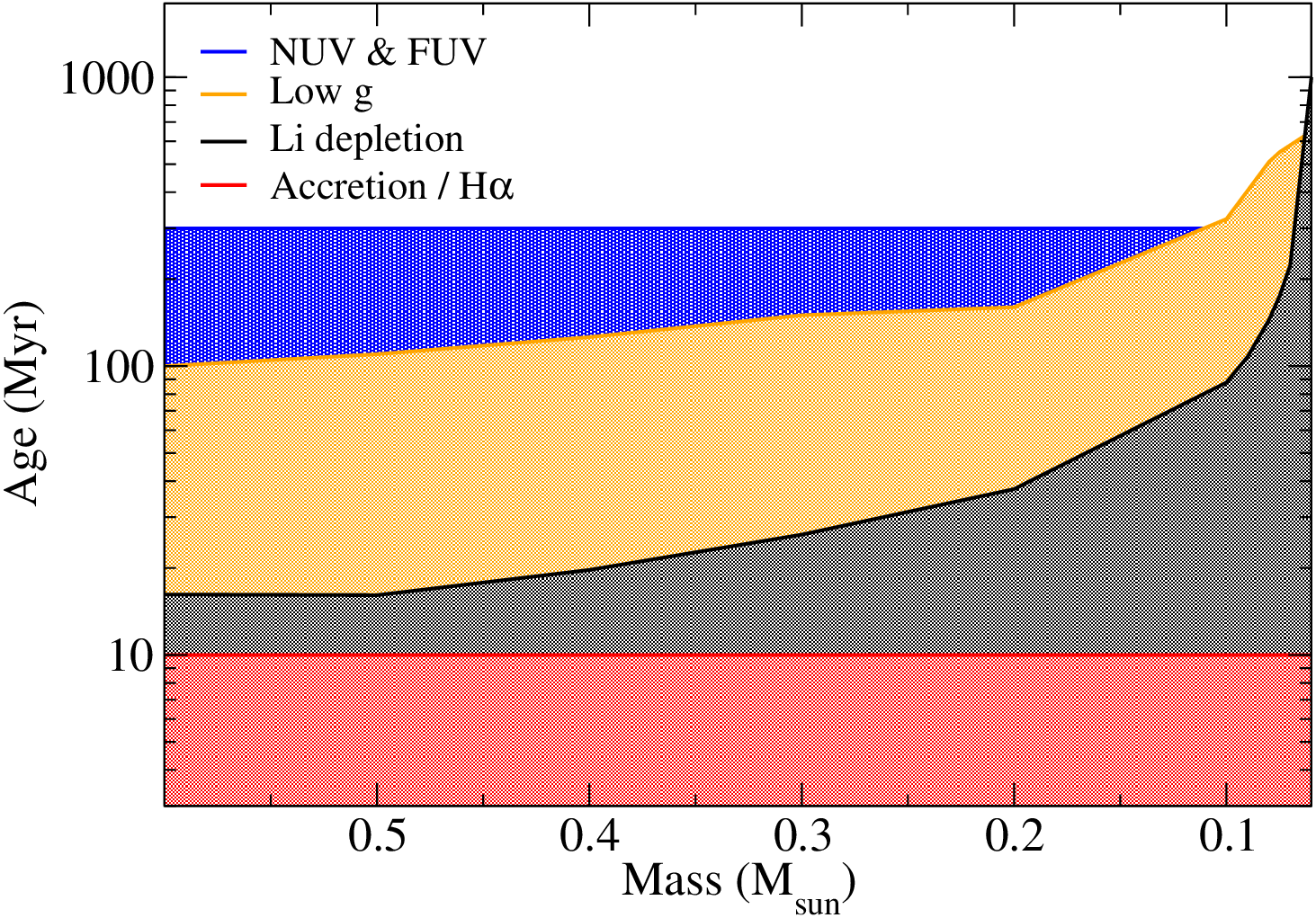}
\caption{A summary of age diagnostics used in this paper.  Each technique provides an upper limit and in the case of lithium, a lower limit if none is detected. The limits set by low-gravity are from evolutionary models of \cite{bara98} and lithium depletion from models of \cite{chab96}. \cite{barr03} set an upper limit of 10 Myr for a star still undergoing accretion. \label{age_diagnostics}}
\end{figure}

Our preliminary selection criterion of strong fractional NUV and FUV flux coincides with the UV emission levels of the strong X-ray emitters of SLR09 and \cite{riaz06} as shown in Figure~\ref{RJ_NUV}. This implies a rough upper limit to the age of our sample of 300 Myr as estimated for the two ROSAT samples.  After identifying the strongest UV emitters in our sample ($F_{FUV}/F_J$ $>$ 10$^{-5}$ and $F_{NUV}/F_J$ $>$
  10$^{-4}$), we used the same spectroscopic age-dating criteria discussed in detail in SLR09 and summarized in Figure~\ref{age_diagnostics}, i.e. low gravity, lithium absorption and strong \ha\ emission as diagnostics of youth.

\subsection{Spectroscopic Youth Indicators}

\begin{figure}
\epsscale{1.05}
\plottwo{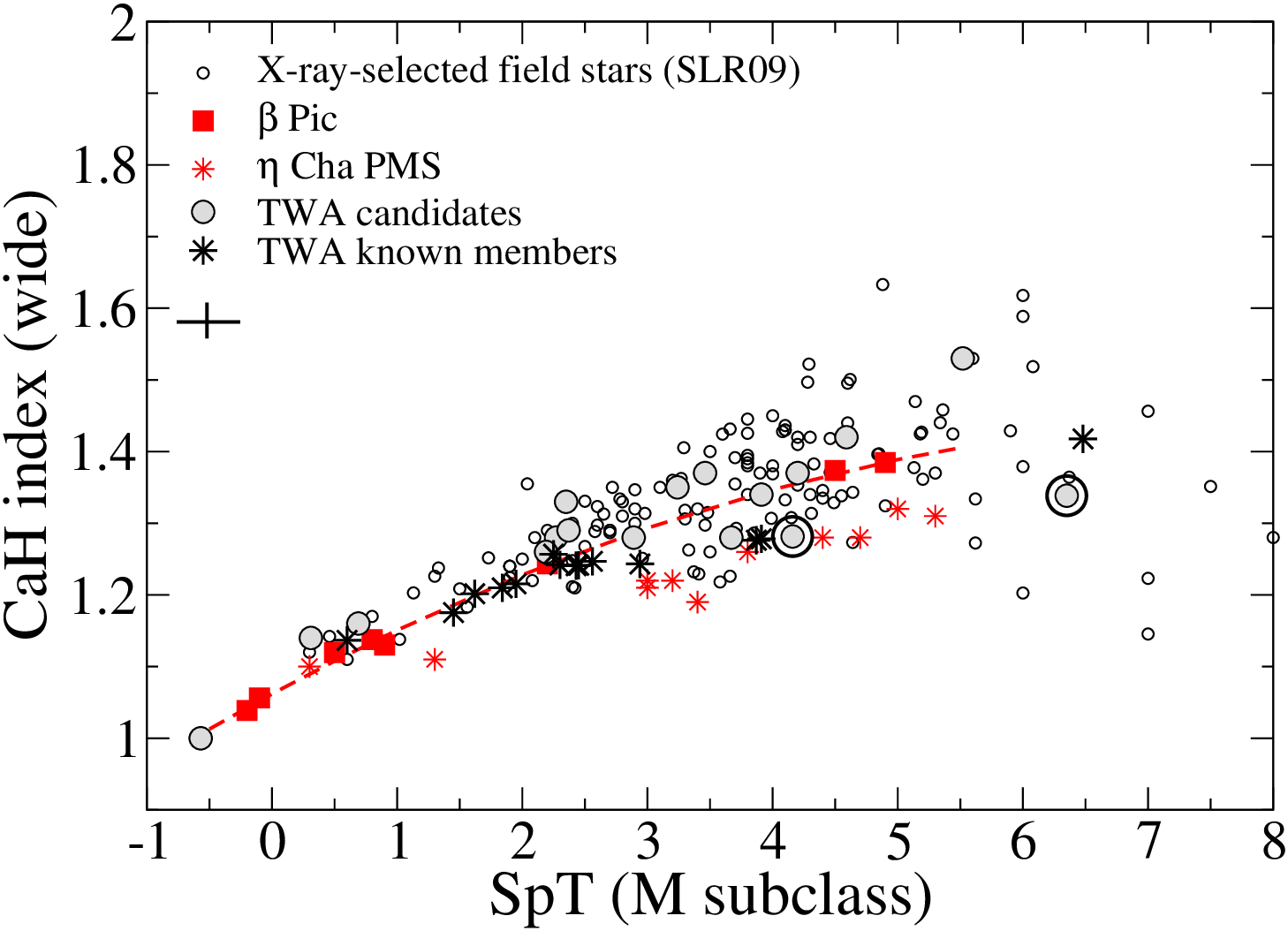}{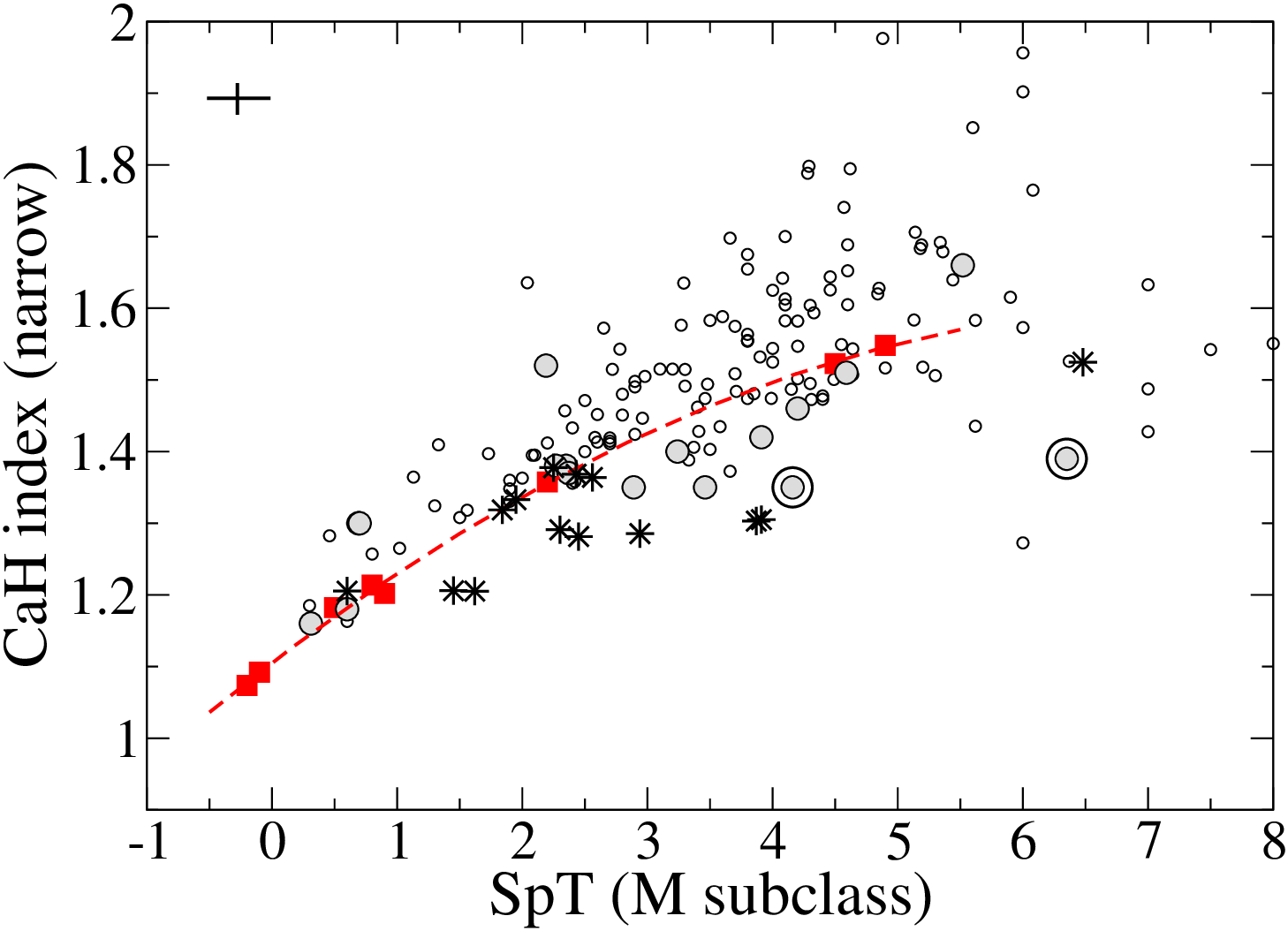}
\caption{The CaH indices (wide on the left; \cite{kirk91}, and narrow on the right; SLR09) for our sample of stars with high resolution optical spectroscopy.  The red dashed curves are polynomial fits to our CFHT and Keck observations of $\beta$ Pic M dwarfs (SLR09): CaH$_{wide}$ = --0.0067 SpT$^2$ + 0.0986 SpT + 1.0633 and CaH$_{narr}$ = --0.00891 SpT$^2$ + 0.13364 SpT + 1.1053. The $\eta$ Cha data are from \cite{lyo04}.\label{SpT_CaH}}
\end{figure}

Models of PMS stars show that lower-mass stars take longer to contract to the main-sequence (MS), e.g.~a 0.5 \Msun~star will reach the MS within 100 Myr whereas a 0.1 \Msun~star will do so in 300 Myr \citep{bara98} and thus determining if a M star has low surface gravity provides an upper limit to its age.
We used the CaH gravity index from \cite{kirk91} defined as the ratio of the mean intensity in two passbands, a `continuum' band and a molecular absorption band of CaH $\lambda$6975: [7020-7050 \AA]/[6960-6990 \AA].
Since we have 15--20 times the resolution of previous M dwarf surveys, we also used a narrower 5-\AA\/ CaH index [7044--7049]/[6972.5--6977.5] providing a more discriminating scale with which to identify low-gravity stars (SLR09). Both indices are plotted as function of SpT in Figure~\ref{SpT_CaH}. 

An important caveat to using the TiO and CaH molecules as temperature and gravity diagnostics is their
dependence on metallicity. Higher metallicity will mimic later spectral types and lower surface gravities. (See discussion in Section 5.1.1 of SLR09.)
We thus also used the K I $\lambda$7699 \AA\ line as a gravity indicator (e.g.~\citealt{sles06}). Care is required with this line as well as it is affected by stellar activity such that higher levels of chromospheric emission fill in the absorption cores and reduce the measured EWs. 

\begin{figure}
\epsscale{.6}
\plotone{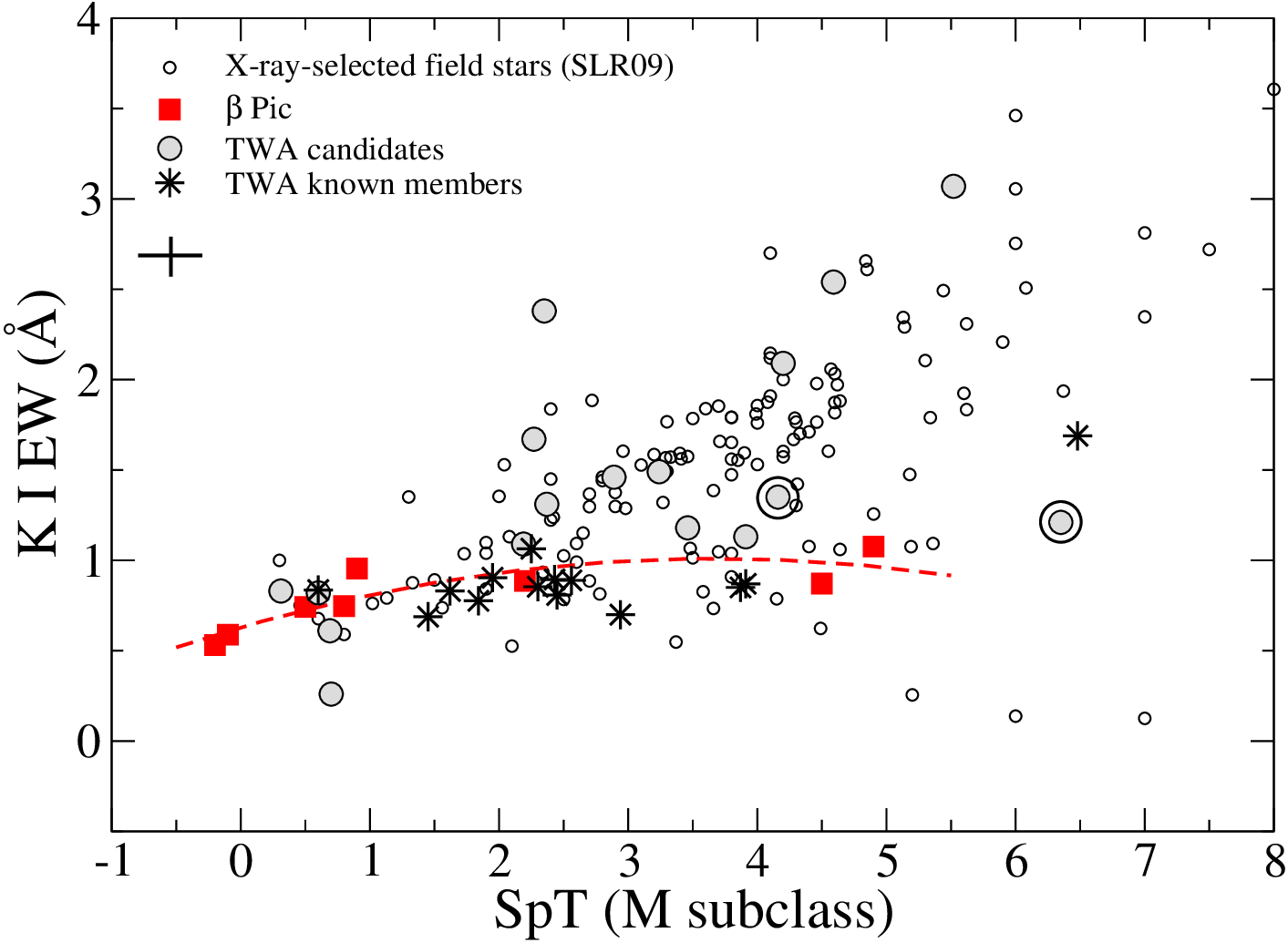}
\caption{K I equivalent widths as a function of spectral type. The red dashed curve is a polynomial fit to the $\beta$ Pic observations from CFHT and Keck (SLR09): EW$_{\rm KI}$ = --0.02821 SpT$^2$ + 0.20731 SpT + 0.62911. \label{SpT_KI}}
\end{figure}

Combining the effects of the chromosphere on K I  with the uncertainties in metallicity on the TiO and CaH indices,
we considered a target as having low-g only if both the CaH {\it and} K I measurements indicate that it is so.  We flagged a target as such in Table~\ref{table_spec} if it falls on or below (within error bars) the best-fit curves to the observed $\beta$ Pic members in Figures~\ref{SpT_CaH} and \ref{SpT_KI}.  Data for the 12-Myr old  $\beta$ Pic members were taken from SLR09.
Out of the 16 UV-bright stars with \ha\ in emission, 8 have low surface gravity with upper age limits of less than 160 Myr \citep{bara98}. Three of these 8 show additional indications of youth and will be discussed in more detail in the following sections.\footnote{The 5 stars with no additional signs of youth have SpTs ranging from M0.3 to M3.9. For these early SpTs, the difference in their predicted masses for a given age produce negligible results in the upper age limits set by the models. If the SpTs were known more precisely, the most significant difference would be in the upper age limit set for the M3.9 star, which would change from 160 Myr to 120 Myr.}

Lower limits on the stellar ages for early M dwarfs are provided for those stars with no lithium absorption ($\lambda$6708 \AA) using the lithium depletion time scales calculated by \cite{chab96}. However, it has been recently shown empirically for at least the $\beta$~Pic moving group, that lower age limits of individual stars based on the lack of lithium absorption systematically over-estimates the star's age as compared to model isochrones \citep{yee10}. In agreement with this, \cite{bara10} recently presented models where stars that are exposed to episodic accretion have higher internal temperatures and thus enhanced lithium depletion compared to stars of the same age and mass but without such accretion. This would imply that the stars with lower age limits in Table~\ref{table_spec} may indeed be younger, perhaps even as young as $\approx$10 Myr.

\begin{figure}
\epsscale{.6}
\plotone{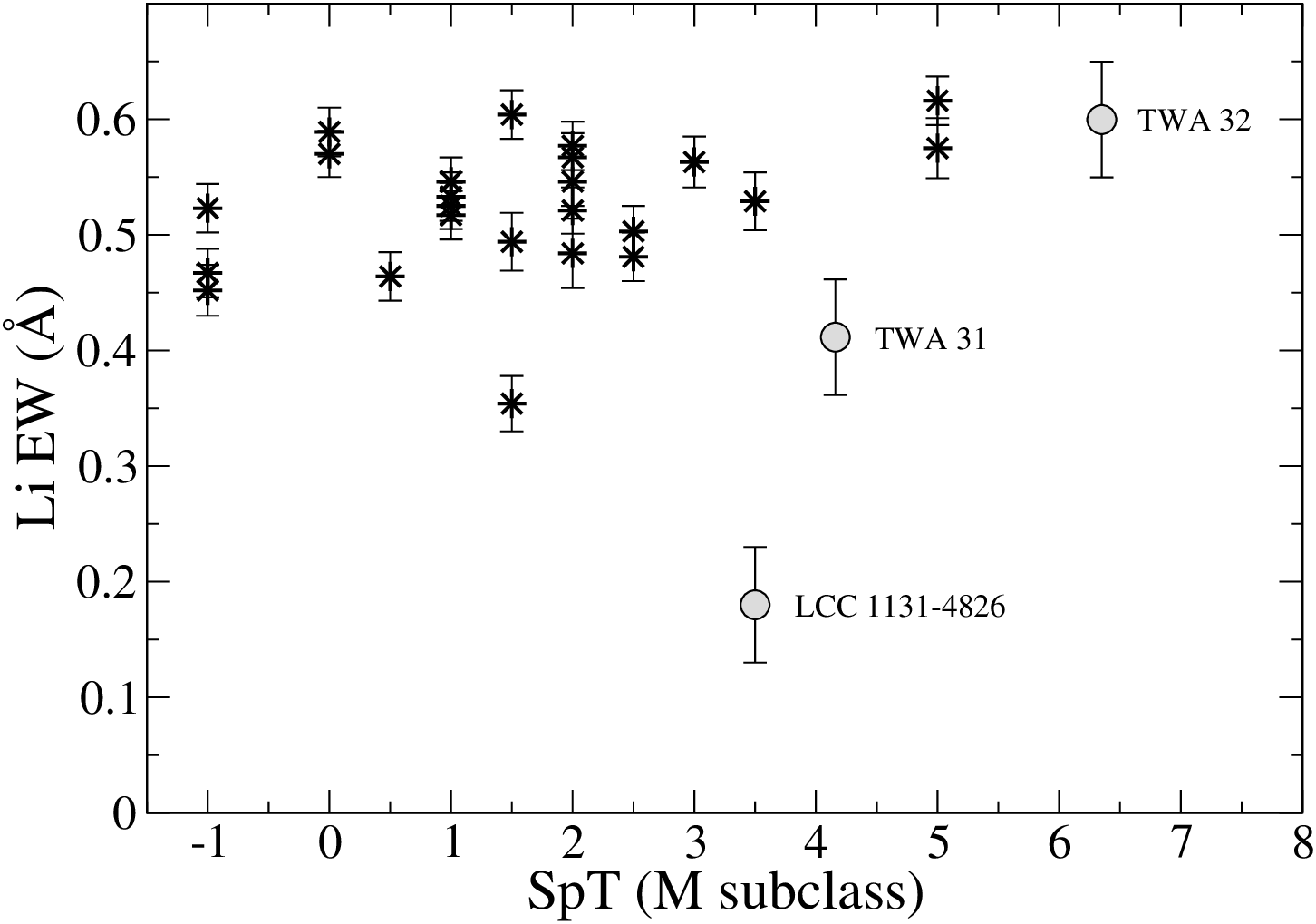}
\caption{Lithium EWs of TWA 31, TWA 32 and LCC 1131-4826 compared with those of known TWA members \citep{ment08}.\label{SpT_Li}}
\end{figure}

We have measured large lithium EWs\footnote{The lithium abundances have not been corrected for possible contamination with the Fe I line at 6707.44 \AA. Uncertainties in the setting of continuum levels prior to measurement induce EW errors of about 0.01--0.02 \AA\ with a dependence on the S/N in the region. We therefore consider our 2$\sigma$ detection limit to be 0.1 \AA.} in 2 of our targets (TWA candidates 1207--3230 and 1226--3316; see Section~\ref{newmembers}) and one marginal detection (0.18 $\pm$ 0.05 \AA) in candidate 1131--4826 (Figure~\ref{SpT_Li}) setting an upper age limit just from the lithium detection of $\approx$40 Myr for the first two and 15 Myr for the last due to its earlier SpT. 

\begin{figure}
\epsscale{.60}
\plotone{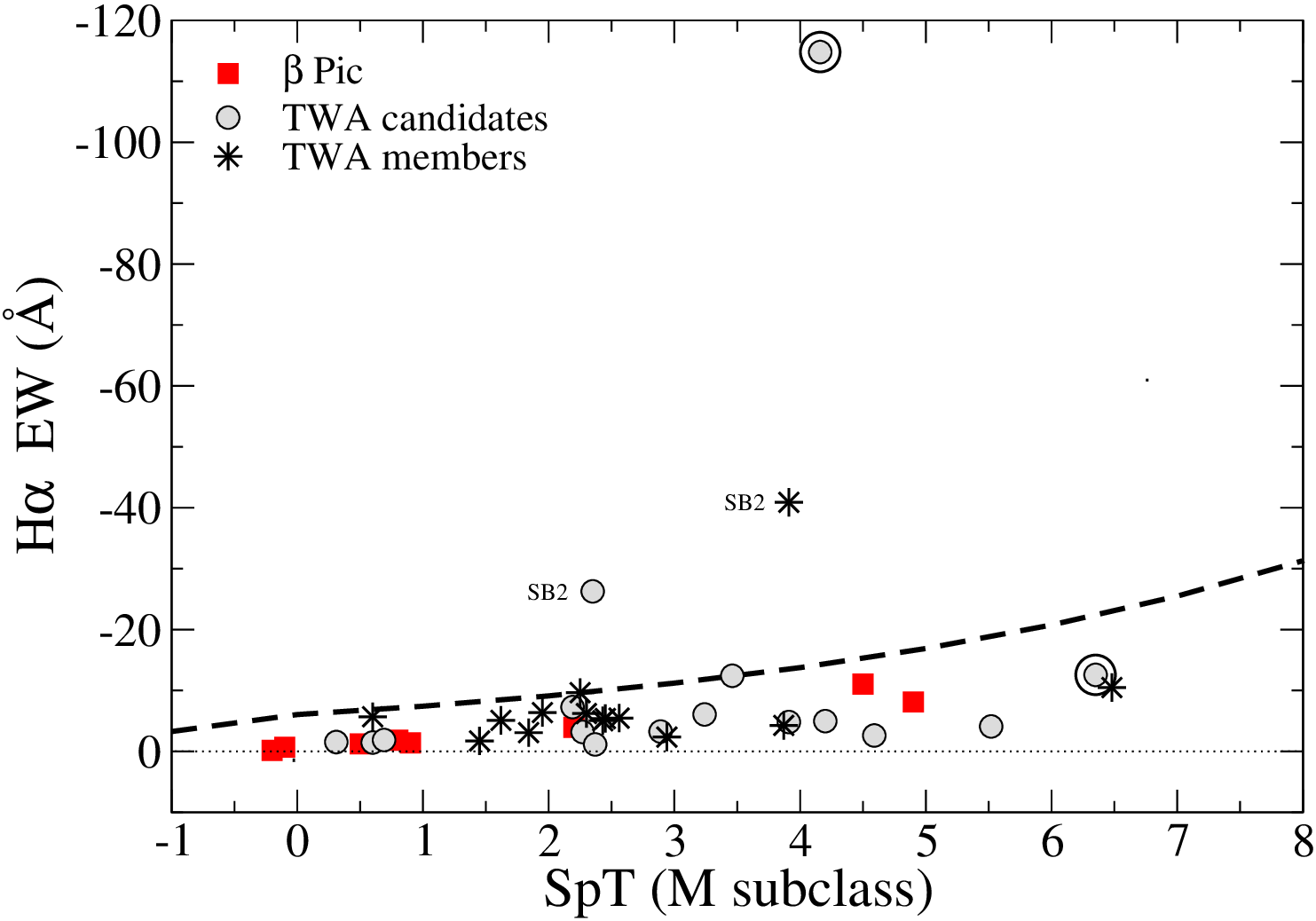}
\caption{\ha\ equivalent widths as a function of spectral type. The dashed curve represents the empirical accretion boundary determined by \cite{barr03}. The 2 SB2s above the accretion curve are TWA 3Aab, which has at least one component still accreting, while the SB2 from our candidate list is not. \label{SpT_Ha}}
\end{figure}

Lastly, the most stringent upper limit is provided by the detection of active stellar accretion. \cite{barr03} have produced an empirical accretion diagnostic by calibrating the \ha\ EW as a function of SpT. This accounts for the increase in EW simply due to the drop in continuum flux from cooler stars (Figure~\ref{SpT_Ha}). Though the accretion curve is not thought to be very robust for objects of SpT later than M5.5, due to the few late-M cluster members used to calibrate the sequence, it does serve as an outer envelope of the chromospheric emission in early Ms. In addition to the EW limits, \cite{whit03} impose a \ha\ 10\% velocity width criterion of $>$270 km~s$^{-1}$ for optically veiled T Tauri stars, and $>$200 km~s$^{-1}$ for non-optically veiled T Tauri stars.

We plot the \ha\ EW of our target stars in Figure~\ref{SpT_Ha}. One of the two strong lithium stars exhibits extremely large \ha\ emission, and is discussed in more detail below.
Two additional stars in Figure~\ref{SpT_Ha} have \ha\ EWs beyond this accretion/non-accretion boundary, both of which are SB2s, including  TWA 3Aab of which at least one component is still accreting as determined by its very broad \ha\ velocity width (395 km~s$^{-1}$). The other SB2, TWA candidate 1013--3542, must have enhanced chromospheric emission due to the tidal spin-up of the two stars, not accretion, as there are no additional spectral signatures of youth, i.e.~no lithium absorption nor any indication of low gravity.

\subsection{Kinematics of the TWA Candidates}\label{kinematics}

\begin{figure}
\epsscale{1}
\plottwo{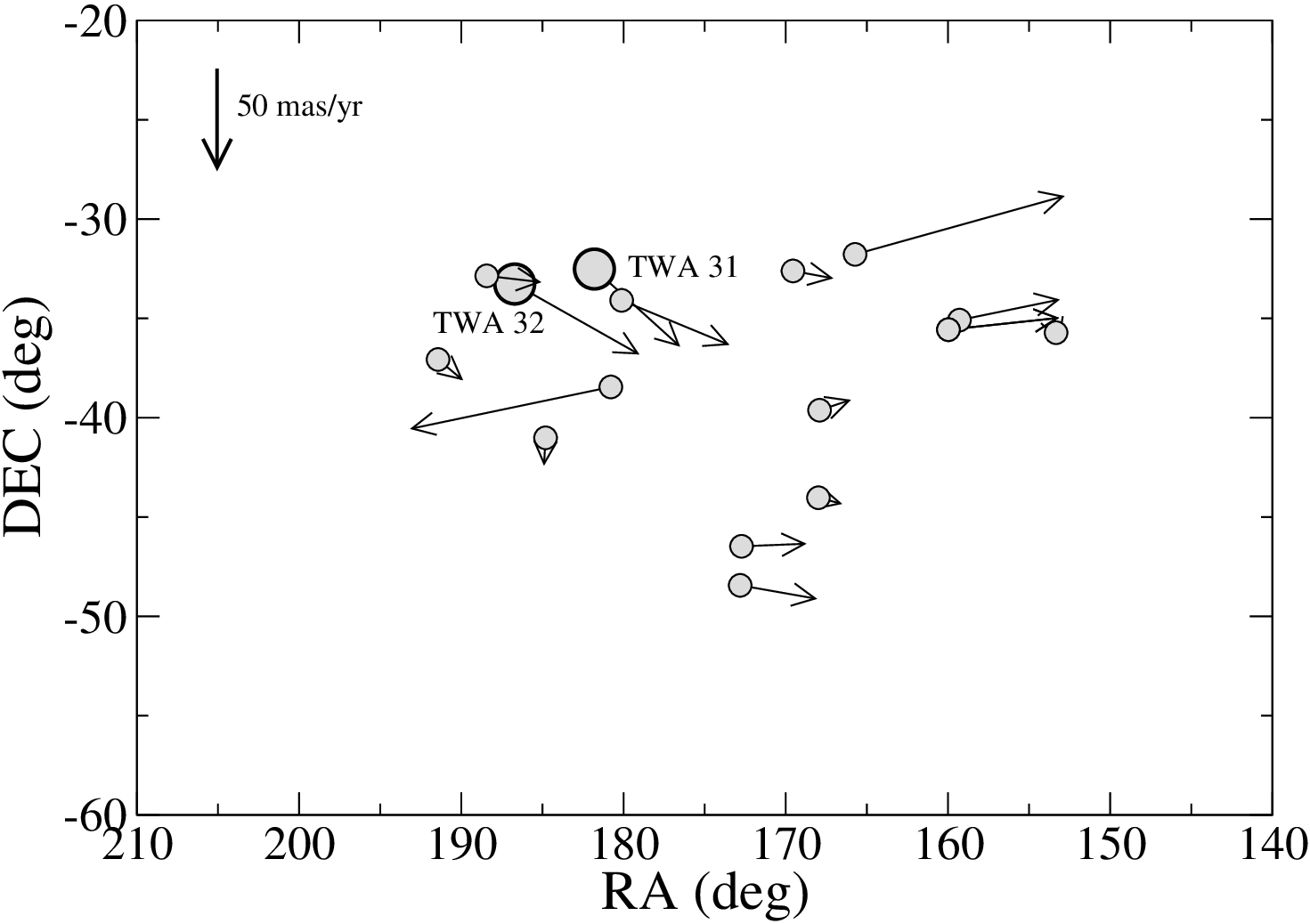}{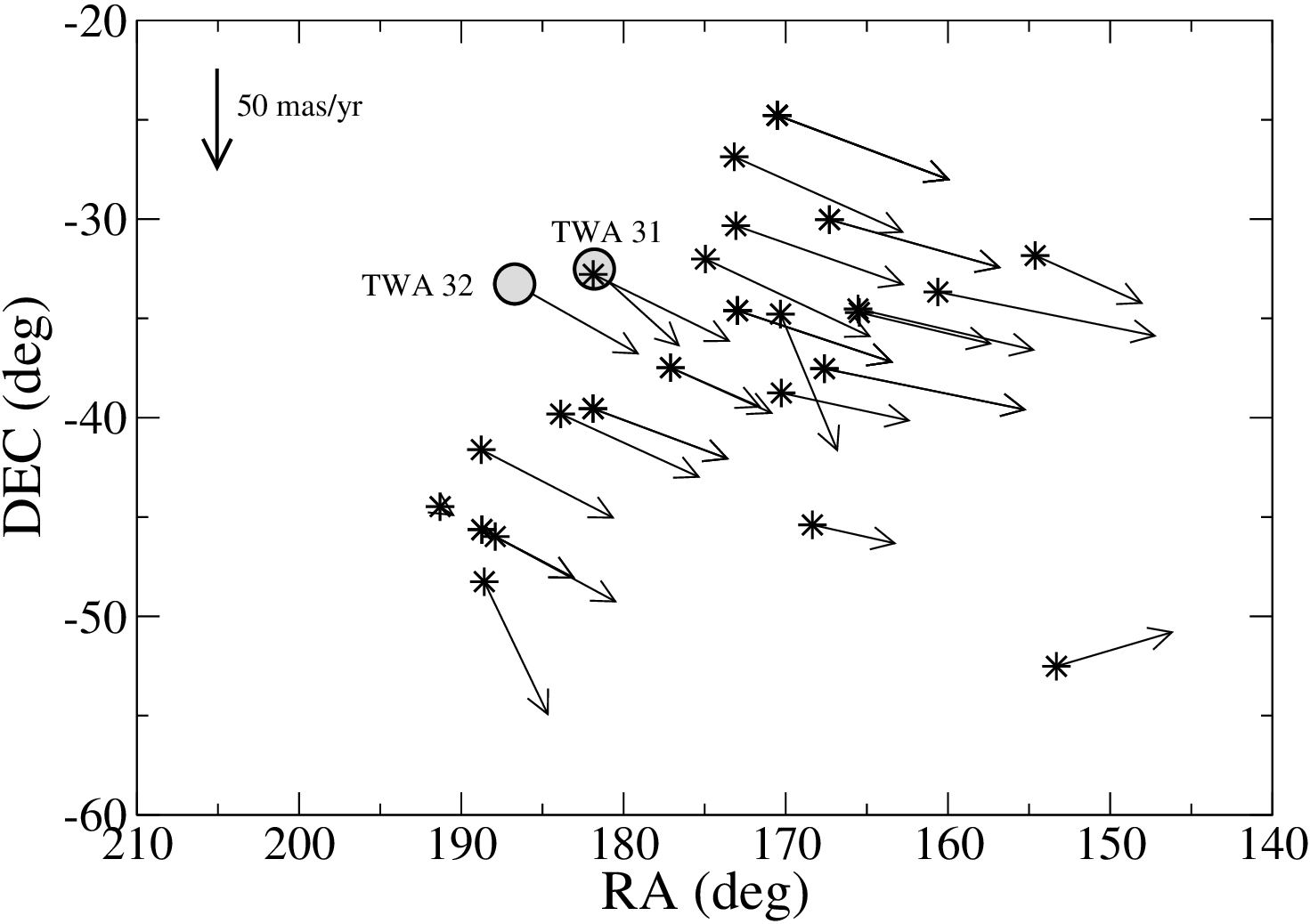}
\caption{RA and DEC positions of candidates (left) and known TWA members (right) with proper motion vectors.  Proper motions with references are listed in Table~\ref{kinematics}.\label{pmradec}}
\end{figure}

The high resolution of the data provided RV measurements to better than 1 km~s$^{-1}$ in almost all cases, which we used in conjunction with the star's photometric distance and proper motions (Figure~\ref{pmradec}) to measure its 3-dimensional space velocity (UVW; \citealt{john87}).  This provides a promising way to determine stellar ages by linking stars kinematically to one of the several known YMGs or associations, including TWA.  

We calculated photometric distances for our sample of TWA candidates, as well as for the known TWA members we observed and have listed them in Tables~\ref{table_twamembers_kinematics} and \ref{table_kinematics}. 
We used the 2MASS $K$ magnitude, $R-K$ or $V-K$ (when available) colors and the age and absolute K magnitudes from the \cite{bara98} models to measure the photometric distance. We took into account the spectroscopically determined ages (or upper age limits) from Table~\ref{table_spec}, since PMS stars are over-luminous compared to dwarfs and will appear closer than they are. For the proposed two new TWA members (discussed below), we adopted the 10 Myr age of TWA \citep{barr06,ment08}. Distances are corrected for any known unresolved binaries. Given the uncertainties in the models, the metallicities of the stars, and the age ranges provided in Table~\ref{table_kinematics}, we estimate the errors to be $\approx$20\% for the non-TWA members and $\approx$10\% for the TWA members, which have more precise ages, and not only upper age limits. 
 
\begin{figure}
\epsscale{1}
\plotone{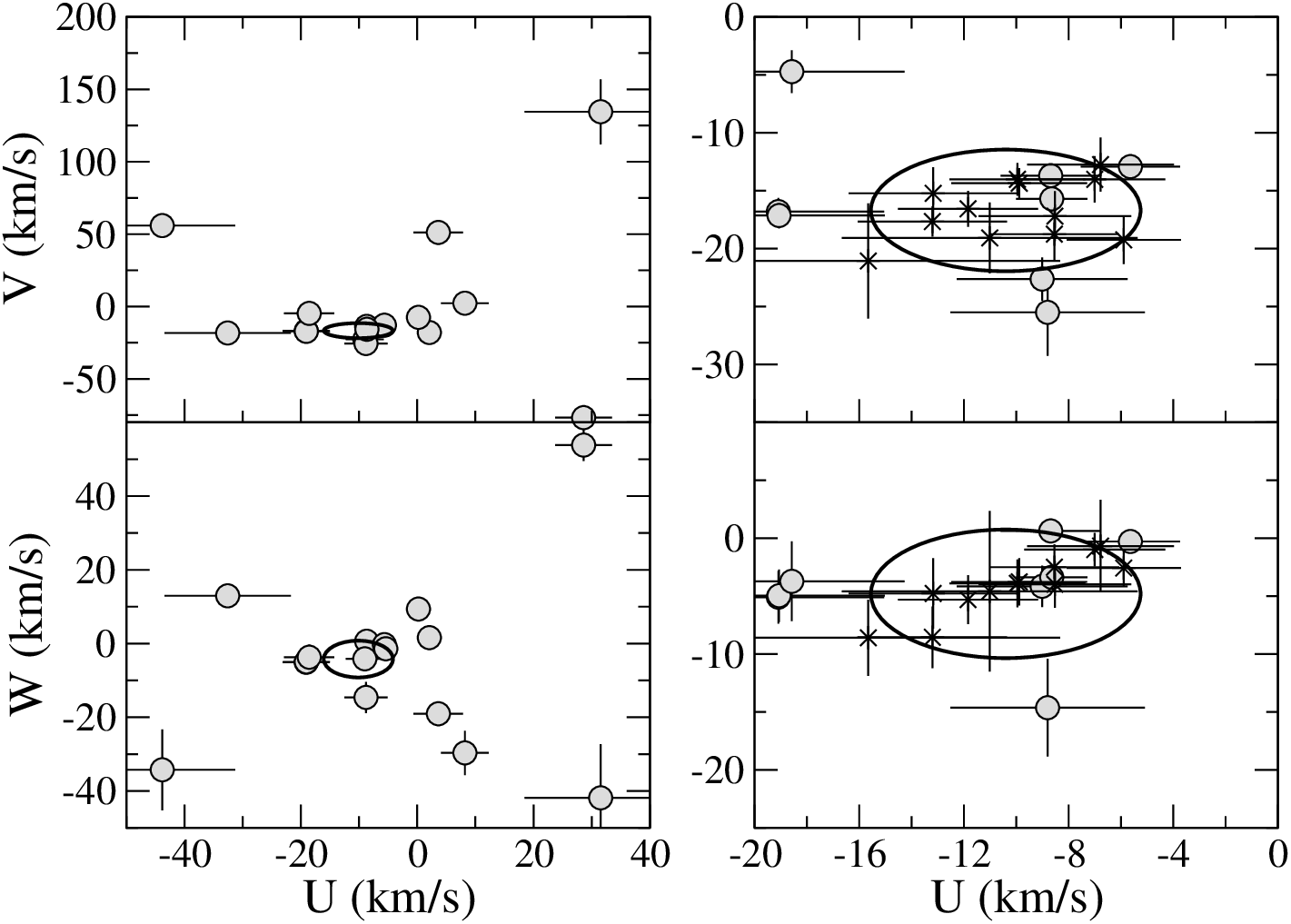}
\caption{Left: UVW velocities of the TWA candidates observed with MIKE.  Right: A zoomed-in view of the 7 candidates clustered around TWA's UVW error ellipse ($\pm2\sigma$) centered on the average UVW=--10.5, --16.9, --4.8 km~s$^{-1}$. UVWs of the known TWA members listed in Table~\ref{table_twamembers_kinematics} are shown as asterisks. \label{twacan_uvw}}
\end{figure}

\begin{figure}
\epsscale{1}
\plotone{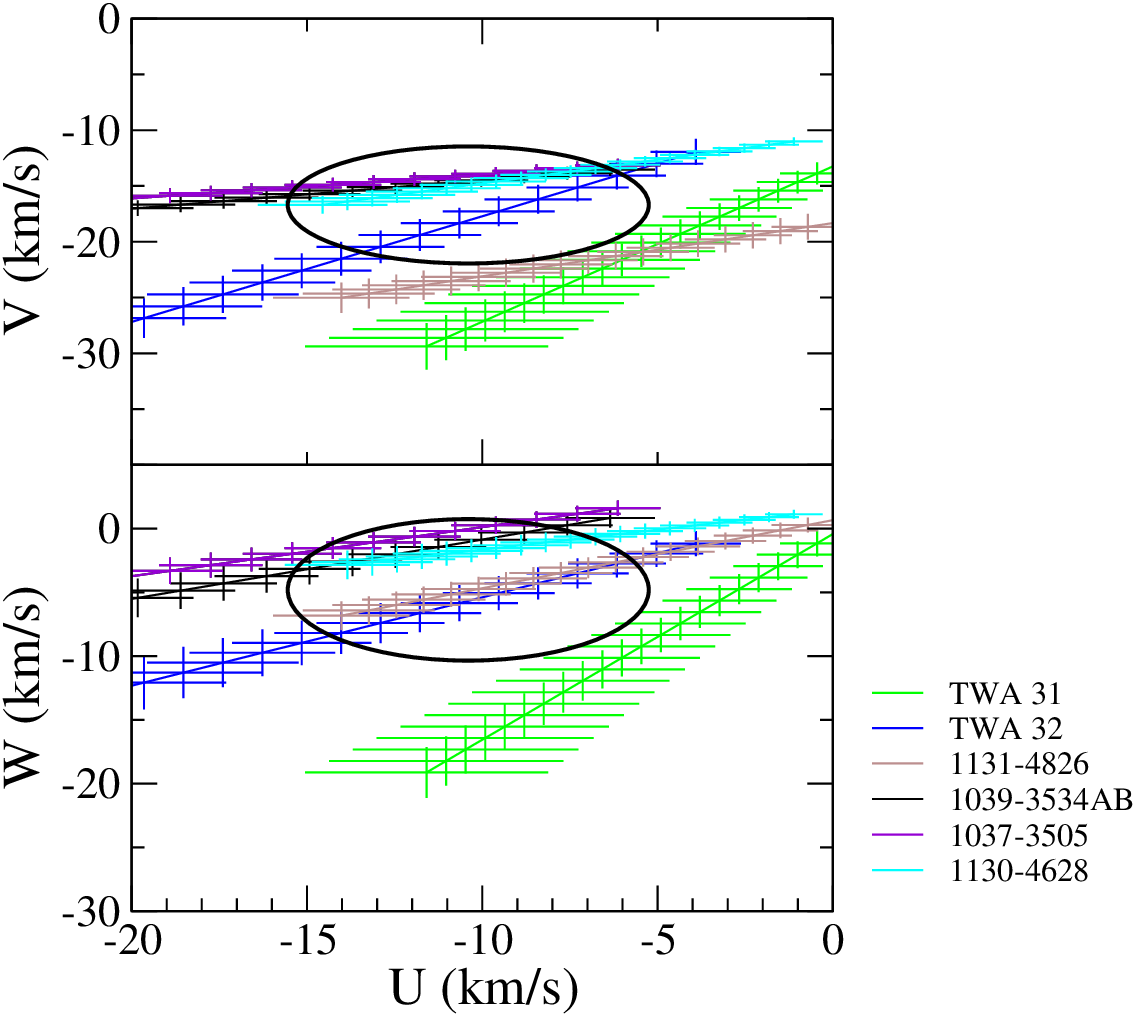}
\caption{UVWs calculated for a range of stellar distances for the 6 TWA candidates that fall near TWA's 2-$\sigma$ error ellipse at their photometric distances. The distances start at 30 pc (top right) and increase by 5 pc increments to a maximum of 130 pc.  \newline \label{twacan_dist_uvw}}
\end{figure}

The UVW velocities for the targets are shown in Figure~\ref{twacan_uvw}.  Seven TWA candidates fall near the 2$\sigma$ error ellipse for the UVWs of known TWA members. Their RA/DEC co-ordinates are 1037--3505, 1039--3534 A, 1039--3534 B, 1130--4628, 1131--4826, 1207--3230, and 1226--3316. And given the large uncertainty in distance, we plot the UVWs of these 7 with a range of possible distances (30--130pc) in Figure~\ref{twacan_dist_uvw}. 

All 7 of the stars appear to have low surface gravity, with only the last two exhibiting strong lithium absorption. (See Section~\ref{newmembers}.) The eighth star listed as low-g in Table~\ref{table_spec}, 2MASS1111-3937, has strong and broad \ha\ emission as well but UVW velocities inconsistent with the ``good UVW box'' defined for young stars by \cite{zuck04}. It is thus probable that 2MASS1111-3937 is an unresolved SB2 with broadened spectral features caught at an orbital phase near conjunction. 

\begin{figure}
\epsscale{1}
\plottwo{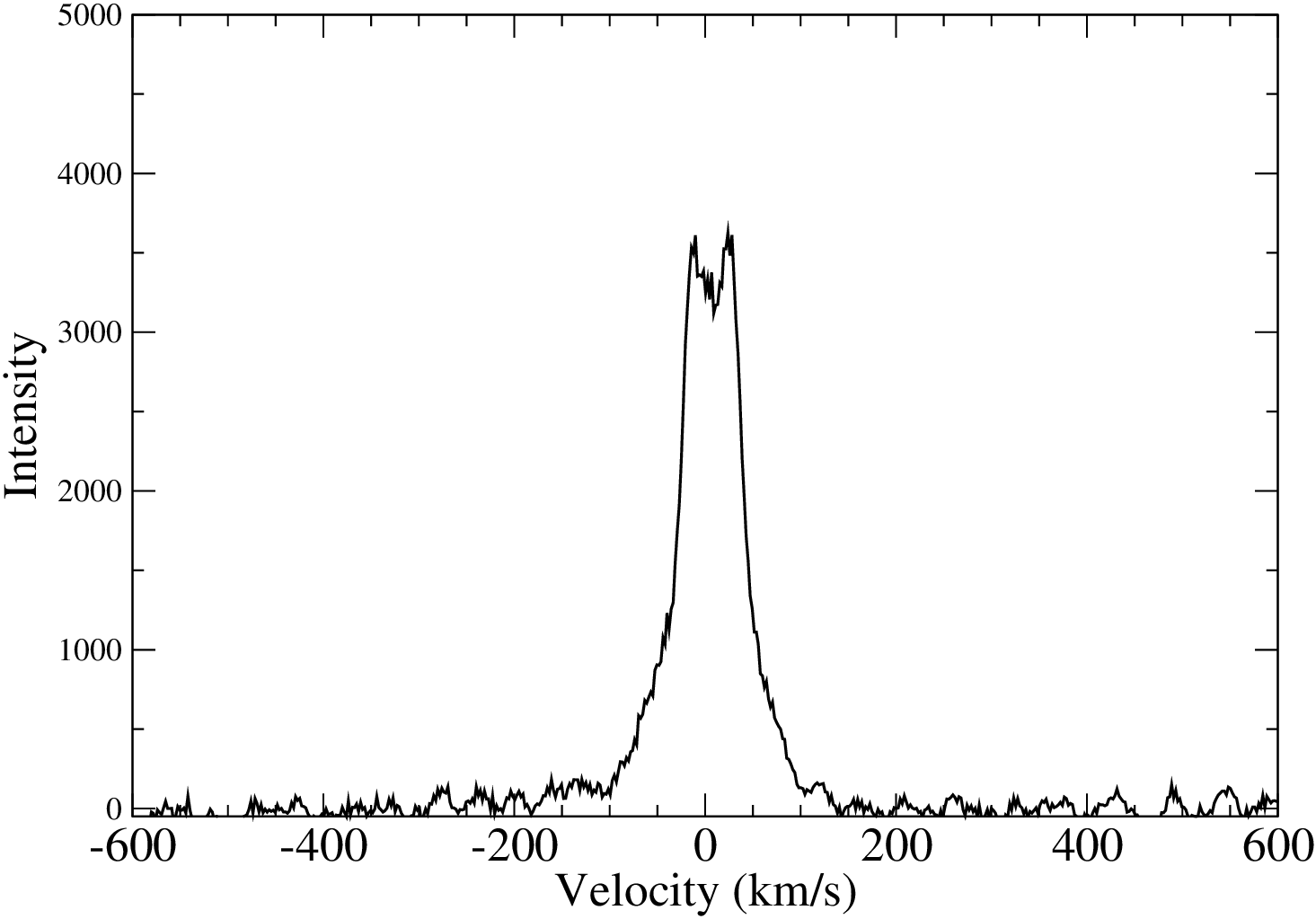}{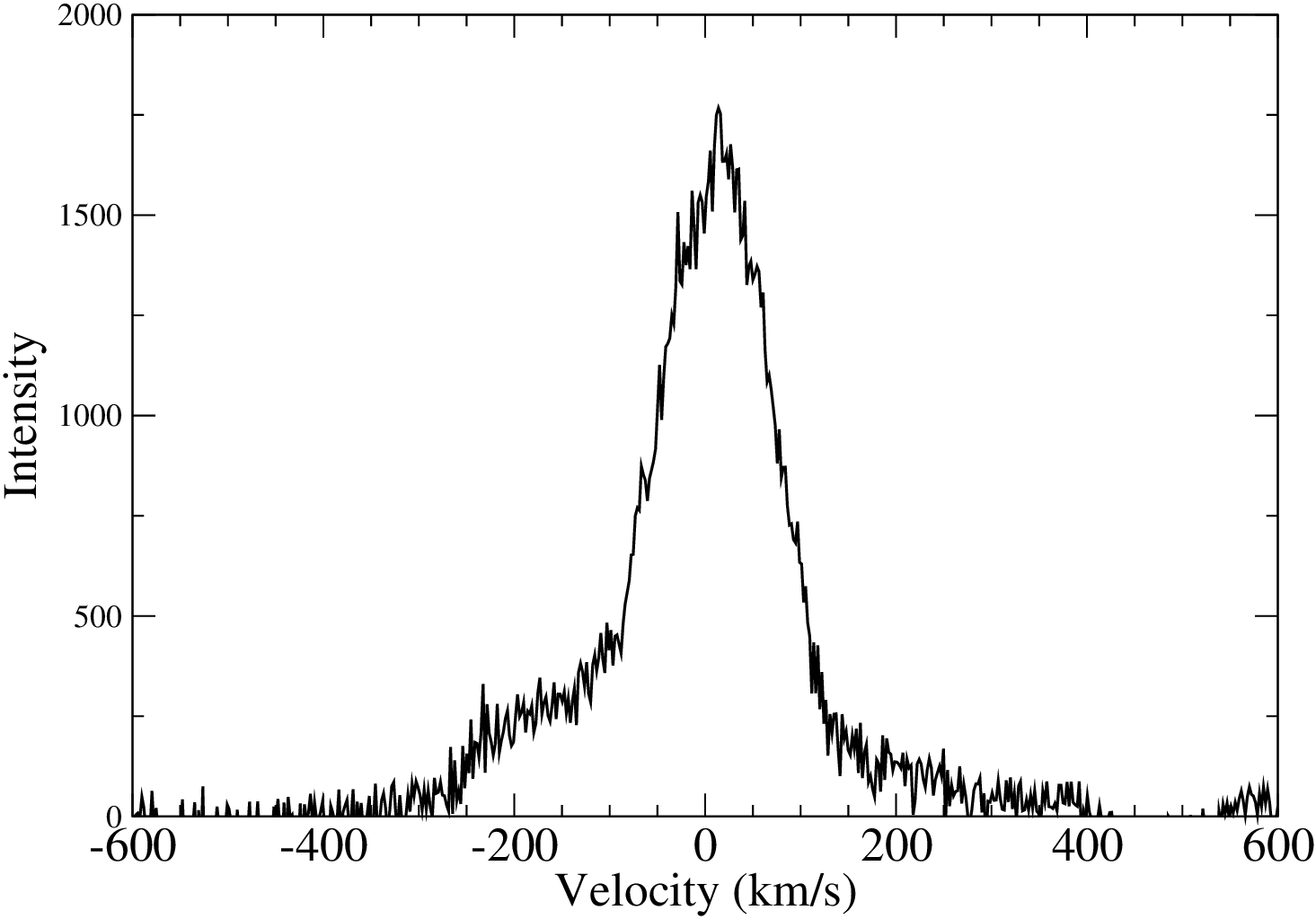}
\caption{\ha\ profiles of TWA 32 (left) and LCC 1131--4826 (right). \label{twa3233_halpha}}
\end{figure}

Candidate 1131--4826 has a weak lithium detection with an EW of 0.18 $\pm$ 0.05 \AA, setting an age limit of 15 Myr. Based on its age, distance of 93 $\pm$ 19 pc, RV of 17.02 $\pm$ 1.16 km~s$^{-1}$ and UVW velocities (--9.0	$\pm$	3.3,	--22.6	$\pm$	1.9,	--4.1	$\pm$	1.8 km~s$^{-1}$), we conclude that this target is a member of LCC \citep{zeeu99,mama02,bitn10} rather than TWA. It is also worth pointing out that LCC 1131--4826 appears to still be accreting based on its broad \ha\ profile (233 km~s$^{-1}$; Figure~\ref{twa3233_halpha}), making it the first known accreting M star in LCC (\citealt{prei08}).

The remaining four low-g stars with UVWs near the TWA UVW error ellipse show no additional signs of youth beyond the UV excess and \ha\ emission, and have upper age limits ranging from 110 -- 300 Myr. They range in distance of 41 -- 84 pc and, assuming lithium is a necessary youth indicator, are not obviously part of either TWA or LCC. 

\section{Two New TWA Members: TWA 31 and TWA 32}\label{newmembers}

Two of our targets share all the same spectroscopic, photometric and kinematic characteristics of known TWA members including low surface gravity, strong Li absorption, strong \ha\/ emission, plus RVs and UVWs consistent with previously known TWA members.  These two likely members have 2MASS coordinates [12:07:10.89~--32:30:53.72] and [12:26:51.35~--33:16:12.47], SpTs of M4.2 and M6.3, and in keeping with tradition, we dubbed them TWA 31 and 32, respectively. They are identified in Tables~\ref{table_phot} and \ref{table_spec}, and are marked by large circles in the figures. Proper motions for the two are from the NOMAD catalog \citep{zach05}, and agree well with the proper motion vectors of known members (Figure~\ref{pmradec}).

\begin{figure}
\epsscale{.60}
\plotone{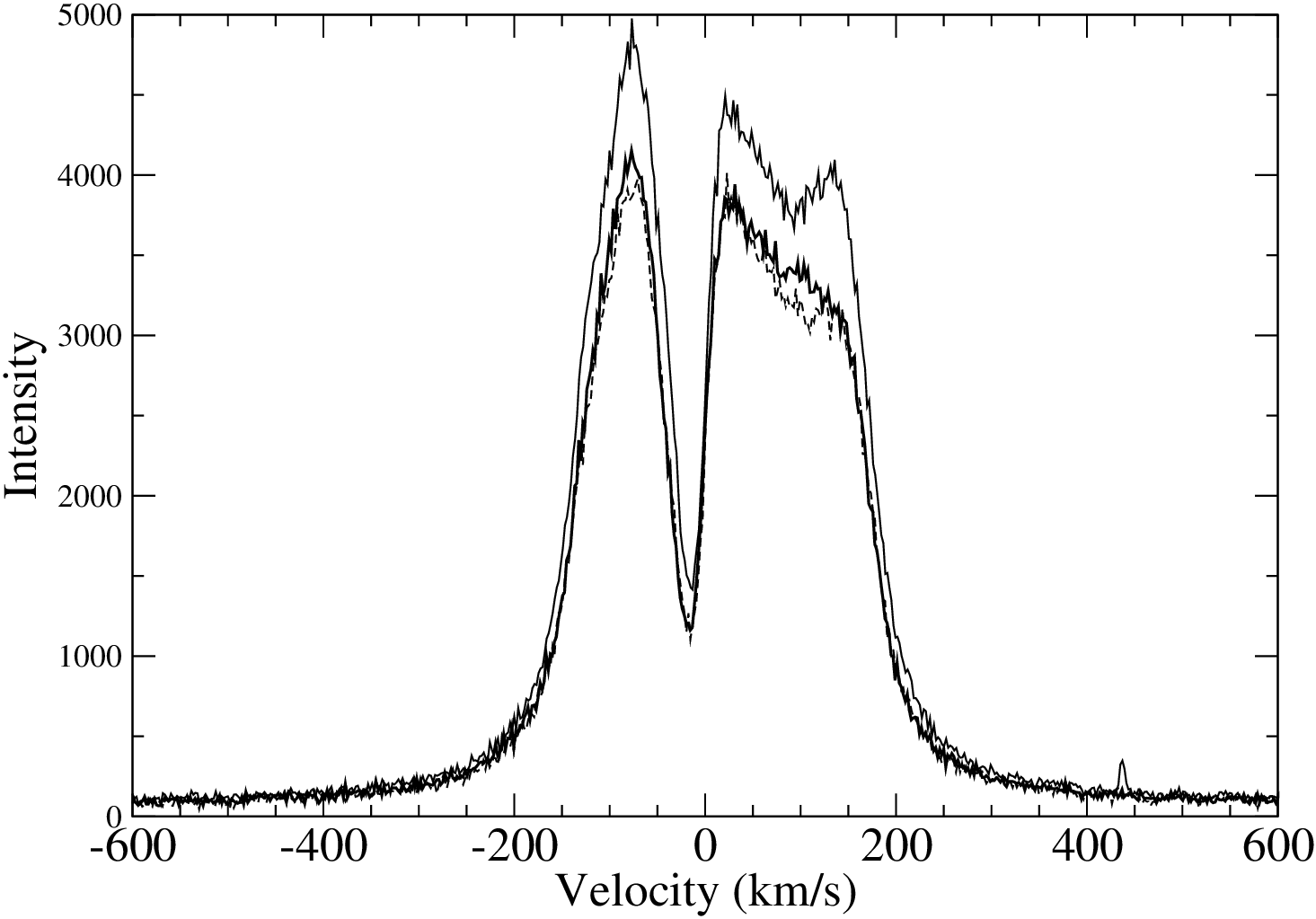}
\caption{\ha\ profiles of TWA 31. Three consecutive 900-s exposures were taken of TWA 31 on UT 2009 June 07. The first observation in the series is the strongest emission profile followed by thick black and thin dashed curves.\label{twa31_halpha}}
\end{figure}

The average Li EW of undisputed members TWA 1--12 from \citep{ment08} is 0.52 with $rms$=0.06 \AA, and TWA 32's EW is consistent with this (0.60 $\pm$ 0.05 \AA).  TWA 31 has a slightly lower-than-average Li EW (0.41 $\pm$ 0.05 \AA), likely due to optical veiling (\citealt{dunc91}). TWA 31 also has by far the strongest \ha\ emission in our sample with an \ha\ EW of --115 \AA\ and an extremely accretion-broadened 10\%-velocity width of 447 km~s$^{-1}$ (Figure~\ref{twa31_halpha}), characteristics comparable to TW Hydrae itself.\footnote{Values for TW Hyd are: log($F_{NUV}/F_J$) = --2.081, log($F_{FUV}/F_J$) = -2.456,  Li EW = 0.467 $\pm$ 0.021 \AA\ \citep{ment08}, \ha\ EW = 220 \AA\ \citep{reid03b}, \ha\ 10\% velocity width = 400 km~s$^{-1}$ \citep{alen02}. } We conclude that TWA 31 is also an accreting T Tauri star with an age of $\lesssim$10 Myr. TWA 31 also emits strongly at He I (EW of --3.6 \AA\ at $\lambda$6678 \AA\ and --10.3 \AA\ at $\lambda$5867 \AA), yet another indication of accretion \citep{moha05}, making it only the 6th known TWA accretor - the other ones being, TW Hyd, Hen 3-600, TWA 14 \citep{muze00b,muze01} and TWA 30A+B \citep{loop10a,loop10b}.

The photometric distances for TWA 31 and 32 are 110 $\pm$ 11 pc and 53 $\pm$ 5 pc (taking binarity into account; see below), respectively.  Although the distance to TWA 31 is relatively large compared to most of the known TWA members, it does not appear to be part of any other neighboring association and is certainly too young to be a member of the 120-pc, 16-Myr old LCC, which is adjacent on the sky to TWA. It is possible that the large distance of TWA 31 implies that it is part of an unidentified young star association in the direction of TWA, as speculated about several other distant, yet clearly young, TWA members (\citealt{wein11}).

\subsection{Keck LGS AO Imaging of TWA 32}

We imaged TWA~32 on UT~2010 May 22 using the sodium laser guide star adaptive optics (LGS AO) system
of the 10-meter Keck II Telescope on Mauna Kea, Hawaii
\citep{2006PASP..118..297W, 2006PASP..118..310V}. We used the facility
IR camera NIRC2 with its 
$10.2\arcsec \times 10.2\arcsec$ field of view during photometric conditions. The LGS provided the wavefront reference source for AO
correction, with the tip-tilt motion measured simultaneously from the
star itself.
We obtained a set of dithered images with the broadband $K$
(2.20~\micron) filter from the Mauna Kea Observatories filter consortium
(\citealt{simo02,toku02}) and easily resolved the
target into a nearly equal-flux binary. Images were reduced in a
standard fashion, and the relative astrometry and photometry were
derived using a multi-gaussian representation of the PSF (e.g.~\citealt{liu08b}). Astrometry was corrected for instrumental distortion, with the
absolute calibration of plate scale and orientation from
\citet{ghez08}. We measured a separation of
$656.1\pm0.4$~mas, a position angle of $11.65\pm0.08$~degs, and a flux
ratio of $0.230\pm0.008$~mag, with the uncertainties determined from the
scatter in the individual images and the overall astrometric calibration
uncertainties.

\section{Summary}\label{summary}

We set out to find the `missing' mid-M dwarfs in TWA by cross-matching optical (HST GSC), infrared (2MASS) and UV (\galex) catalogs, filling in the stellar mass function in the association and providing excellent targets for direct imaging searches for substellar companions and circumstellar disks. \galex\ provides a new and more sensitive resource that enables a major expansion of the young low-mass census, far beyond previous data sets. We found that NUV data yield many
candidates, but applying FUV criteria provide an excellent means to distinguish between the (never-before-delineated) quiescent
emission of old stars ($F_{FUV}/F_J$ $<$ 10$^{-5}$), the faint sources (FUV not detected), and the truly young targets
with high levels of UV emission ($F_{FUV}/F_J$ $>$ 10$^{-5}$ and $F_{NUV}/F_J$ $>$ 10$^{-4}$ for $R-J \gtrsim 1.5$).
The photometric cross-matching yielded 34 UV-bright low-mass stars with
SpT between M0 and M6 within $\approx$100 pc and 1000 sq.~deg.~of the TWA with ages likely less than 300 Myr.

Ground-based optical low-resolution spectroscopy of 30 identified 16 with \ha\ emission which were followed up with high-resolution spectroscopy.\footnote{Several of the UV bright targets with no \ha\ emission are white dwarf+M dwarf pairs.} Of these, 2 (possibly 3) are old SB2s with tidally enhanced UV emission. 
Six are nearby field Ms with ages probably younger than 300 Myr, based on their strong UV emission with no additional signs of youth.  
One candidate appears to be an accreting new member of the 16-Myr old LCC, and 5 are low-gravity M dwarfs with maximum ages ranging from 110 -- 300 Myr.  Four of these 5 low-g stars are kinematically identical to the previously known TWA members, yet are likely older than 20 Myr based on the absence of Li absorption. Thus identifying new YMG members based on kinematics and strong X-ray or UV emission alone may not be sufficient and spectroscopic observations are necessary for confirmation. However, in light of the recent work by \cite{bara10}, these stars may be much younger and possibly TWA members despite not having any Li absorption.

Lastly, 2 stars in our sample exhibit all the spectroscopic, photometric and kinematic characteristics of $\approx$10 Myr-old TWA
members including low surface gravity, strong Li absorption, strong \ha\ emission and UVW velocities.
These new members, TWA 31 (SpT=M4.2) and TWA 32 (SpT=M6.3), have photometric distance of 110 $\pm$ 11 pc and 53 $\pm$ 5 pc, respectively.
Followup Keck/LGS AO observations resolved TWA 32 into two near equal-flux ($0.230\pm0.008$~mag in K) stars with a separation of 656.1
$\pm$ 0.4 mas. 
TWA 31 also exhibits an extremely
accretion-broadened \ha\ profile (447 $\pm$ 10 km~s$^{-1}$) with a slightly lower-than-average Li EW (0.41 $\pm$ 0.05 \AA), likely due to
optical veiling (\citealt{dunc91}), making it only the 6th known active accretor in TWA.

Our new \galex/AIS search method successfully recovered 2/3 of the known TWA members (corresponding to the 2/3 of the sky covered in the GR4 data archive release), making it surprising that only 2 new mid-M members were discovered. With the peak of the spectral-type (and mass) function at SpT=M3--M4 (\citealt{boch10} and references therein), the expected number of newly found M dwarfs is substantially greater than 3, likely closer to 20 (based on the known number of early-Ms in TWA). Our results imply that either TWA has an unexpected mass function, or a significant fraction of 10-Myr M dwarfs have depleted all their lithium and were eliminated from the membership list.  This latter possibility would imply that low-g stars that are kinematically identical to TWA but lacking Li may indeed be bona fide members of the association.

\acknowledgements

E.S. thanks Bernie Shaio and Tony Rogers of MAST/STScI for \galex/AIS query support. Also, we appreciate the helpful comments on the manuscript by the referee, Eric Mamajek and useful discussions with Andrew West and John Debes. This material is based upon work
supported by the Carnegie Institution of Washington and the NASA/\galex\ grant program under Cooperative Agreement Nos. NNA04CC08A and
NNX07AJ43G issued through the Office of Space Science. This publication makes use of data products from the \galex\ All-sky Imaging
Survey, the HST Guide Star Catalog (v2.3) and the Two Micron All Sky Survey, with access to the last two provided by Vizier and SIMBAD.
2MASS is a joint project of the University of Massachusetts and the Infrared Processing and Analysis Center/California Institute of
Technology, funded by the National Aeronautics and Space Administration and the National Science Foundation.

\clearpage
\bibliography{refs}{}
\bibliographystyle{apj}

\clearpage 
\begin{deluxetable}{lcccccccccccc}
\tabletypesize{\scriptsize}
\rotate
\tablecaption{TWA Candidate M dwarfs Observed with MagE-- Photometric Data\label{table_phot}}
\tablewidth{0pt}
\tablehead{	
\colhead{RA \& DEC} & 		 \colhead{$l$} & 		\colhead{$b$} &  \colhead{$R$} & 		\colhead{$K$} & 			\colhead{$J-H$} &  \colhead{$H-K$} 	 & \colhead{log$(F_{NUV}/F_J)$}\tablenotemark{a} & \colhead{log$(F_{FUV}/F_J)$}\tablenotemark{a} & \colhead{\ha\ } & \colhead{Note}\\
\colhead{{$_{2MASS}$}} & \colhead{deg} & \colhead{deg} & \colhead{$_{GSC2.3}$} & \colhead{$_{2MASS}$} &  \colhead{} & 	\colhead{} & 		\colhead{} & \colhead{}  & \colhead{emission?} & \colhead{}
}
\startdata
							
10	06	24.73	-30	14	43.98	&	265.489	&	20.388	&	14.59	&	10.593	&	0.537	&	0.275	&	-3.80	&	-4.08	&	no	&		\\
10	13	21.43	-35	42	36.93	&	270.235	&	16.942	&	17.38	&	13.510	&	0.575	&	0.307	&	-1.33	&	-1.31	&	yes	&		\\
10	30	54.53	-40	11	27.58	&	275.832	&	15.176	&	11.97	&	7.622	&	0.941	&	0.287	&	-3.76	&	-3.81	&	no	&		\\
10	37	10.47	-35	05	01.52	&	274.116	&	20.167	&	14.69	&	10.803	&	0.555	&	0.310	&	-3.29	&	-3.69	&	yes	&		\\
10	39	52.76	-35	34	03.03	&	274.886	&	20.040	&	11.67	&	9.038	&	0.582	&	0.160	&	-3.58	&	-4.14	&	yes	&	VB, \rosat\ source	\\
11	02	53.73	-31	45	10.57	&	277.441	&	25.691	&	17.61	&	13.687	&	0.493	&	0.301	&	-1.81	&	-2.51	&	yes	&		\\
11	06	02.46	-31	05	08.38	&	277.778	&	26.590	&	18.39	&	13.710	&	0.482	&	0.327	&	-0.53	&	-2.46	&	no	&		\\
11	11	46.37	-39	37	34.78	&	282.820	&	19.321	&	18.17	&	13.799	&	0.711	&	0.268	&	-1.53	&	-1.76	&	yes	&		\\
11	11	52.67	-44	01	53.87	&	284.642	&	15.294	&	15.50	&	11.223	&	0.593	&	0.269	&	-2.87	&	-3.39	&	yes	&		\\
11	18	12.37	-32	35	59.09	&	281.095	&	26.304	&	18.79	&	14.075	&	0.584	&	0.313	&	-2.93	&	-3.01	&	yes	&		\\
11	30	53.55	-46	28	25.19	&	288.778	&	14.173	&	14.84	&	11.286	&	0.522	&	0.283	&	-3.05	&	-3.38	&	yes	&		\\
11	31	14.83	-48	26	27.98	&	289.463	&	12.329	&	13.14	&	9.605	&	0.617	&	0.256	&	-3.25	&	-3.85	&	yes	&	LCC member	\\
11	34	56.51	-36	40	28.09	&	286.257	&	23.716	&	19.18	&	14.262	&	0.595	&	0.395	&	-1.35	&	-1.23	&	no	&		\\
11	34	58.90	-34	43	11.59	&	285.575	&	25.563	&	15.67	&	11.921	&	0.534	&	0.264	&	-3.00	&	-3.00	&	no	&		\\
11	39	08.06	-45	32	39.81	&	289.896	&	15.501	&	16.15	&	12.313	&	0.584	&	0.261	&	-1.50	&	-2.87	&	no	&		\\
11	45	50.03	-33	02	13.32	&	287.404	&	27.849	&	12.55	&	9.356	&	0.193	&	0.068	&	-1.53	&	-3.62	&	no	&	CD-32 8299	\\
11	48	08.96	-37	58	09.45	&	289.406	&	23.236	&	18.57	&	14.213	&	0.621	&	0.345	&	-2.13	&	-1.93	&	no	&		\\
12	00	27.51	-34	05	37.17	&	291.035	&	27.598	&	13.35	&	8.723	&	0.622	&	0.261	&	-3.86	&	-4.64	&	no	&		\\
12	03	08.07	-38	26	55.54	&	292.643	&	23.465	&	11.66	&	8.565	&	0.594	&	0.271	&	-3.77	&	-4.54	&	yes	&	\rosat\ source 32\arcsec away	\\
12	07	10.89	-32	30	53.72	&	292.208	&	29.457	&	16.56	&	12.115	&	0.558	&	0.375	&	-2.89	&	-2.81	&	yes	&	TWA 31 	\\
12	19	07.68	-41	01	57.81	&	296.393	&	21.441	&	16.93	&	13.012	&	0.566	&	0.303	&	-1.74	&	-2.42	&	yes	&		\\
12	19	29.82	-34	14	24.54	&	295.447	&	28.178	&	19.07	&	14.347	&	0.711	&	0.331	&	-1.69	&	-2.15	&	no	&		\\
12	20	25.78	-43	04	06.97	&	296.934	&	19.442	&	17.97	&	13.719	&	0.589	&	0.367	&	-1.99	&	-2.12	&	no	&		\\
12	21	33.27	-41	40	29.12	&	296.968	&	20.865	&	17.97	&	13.866	&	0.706	&	0.261	&	-1.43	&	-1.35	&	no	&		\\
12	24	08.36	-31	34	27.48	&	296.156	&	30.940	&	18.11	&	13.850	&	0.624	&	0.250	&	-2.57	&	-2.50	&	no	&		\\
12	26	43.75	-41	47	37.31	&	298.004	&	20.856	&	12.54	&	9.082	&	0.817	&	0.267	&	-1.55	&	-3.63	&	no	&		\\
12	26	51.35	-33	16	12.47	&	297.040	&	29.320	&	14.92	&	9.783	&	0.569	&	0.339	&	-3.38	&	-3.32	&	yes	&	TWA 32 	\\
12	33	43.57	-32	51	26.29	&	298.636	&	29.877	&	16.95	&	12.149	&	0.556	&	0.317	&	-2.62	&	-2.41	&	yes	&		\\
12	45	43.60	-37	04	36.24	&	301.673	&	25.790	&	18.85	&	13.649	&	0.556	&	0.398	&	-2.14	&	-2.40	&	yes	&		\\
12	56	49.64	-30	07	37.37	&	304.317	&	32.730	&	16.95	&	12.883	&	0.655	&	0.294	&	-2.60	&	-2.82	&	no	&		\\
\enddata
\tablenotetext{a}{Note that $F_J$ is calculated using the full width of the 2MASS J band filter, 0.3 \micron.}

\end{deluxetable}

\begin{deluxetable}{lccccccccccrrlllll}
\tabletypesize{\scriptsize}
\rotate
\tablecaption{TWA members observed with MIKE\label{twa_members}}
\tablewidth{0pt}
\tablehead{
\colhead{Name}   & \colhead{log$(F_{NUV}/F_J)$} & \colhead{log$(F_{FUV}/F_J)$} & \colhead{SpT}  &  \colhead{CaH-narr.} & \colhead{K I EW} & \colhead{Li EW} & \colhead{H$\alpha$ EW}   & \colhead{Binarity}\\
\colhead{} &  \colhead{} & \colhead{} & \colhead{M--} & \colhead{Index} & \colhead{\AA} & \colhead{\AA}  & \colhead{\AA} &  \colhead{}\\
\colhead{} & \colhead{} & \colhead{} & \colhead{($\pm 0.5$)}  &  \colhead{($\pm 0.03$)} & \colhead{($\pm 0.05$)} & \colhead{($\pm 0.2$)} & \colhead{($\pm 0.05$)} & \colhead{} & \colhead{}

}
\startdata
TWA 2AB	&	-3.87	&	-4.62	&	1.5	&	1.21	&	0.69	&	0.52	&	-1.72	&	VB \citep{bran03}	\\
TWA 3Aab	&	-3.194\tablenotemark{a}	&	-3.64	&	3.9	&	1.31	&	0.87	&	0.51	&	-40.89	&	VB, SB2 \citep{muze00}	\\
TWA 3B	&	-3.194\tablenotemark{a}	&	-3.64	&	3.9	&	1.30	&	0.85	&	0.54	&	-4.26	&	VB	\\
TWA 5Aab	&	-3.57	&	-4.19	&	1.9	&	1.33	&	0.90	&	0.63	&	-6.37	&	VB, SB2 \citep{torr06}	\\
TWA 7	&	not observed	&		&	2.4	&	1.28	&	0.81	&	0.55	&	-5.39	&		\\
TWA 8A	&	-3.61	&	-4.36	&	2.4	&	1.37	&	0.89	&	0.55	&	-5.04	&	VB	\\
TWA 8B	&	-3.72	&	--	&	5\tablenotemark{b}	&	1.29	&	0.85	&	0.58	&	-6.21	&	VB	\\
TWA 10	&	-3.87	&	-4.50	&	2.6	&	1.36	&	0.89	&	0.50	&	-5.46	&		\\
TWA 12	&	-3.72	&	-4.39	&	1.6	&	1.21	&	0.83	&	0.53	&	-5.10	&		\\
TWA 14ab	&	-2.90	&	-3.42	&	0.6	&	1.21	&	0.84	&	0.59	&	-5.68	&	SB2 \citep{jaya06}	\\
TWA 15B	&	not observed	&		&	2.2	&	1.38	&	1.06	&	0.55	&	-9.64	&	VB	\\
TWA 16	&	-3.81	&	-4.18	&	1.8	&	1.32	&	0.78	&	0.38	&	-3.08	&		\\
TWA 22AB	&	not observed	&		&	6.5	&	1.52	&	1.69	&	0.65	&	-10.48	&	VB \citep{bonn09}	\\
TWA 23ab	&	-4.05	&	-4.65	&	2.9\tablenotemark{c}	&	1.29	&	0.70	&	0.50	&	8.12	&	SB2 (this work)	\\
\enddata

\tablenotetext{a}{\galex\ cannot resolve TWA 3A and 3B.}
\tablenotetext{b}{The TiO index gave SpT=M3, but we list the published SpT from \cite{torr03}.}
\tablenotetext{c}{Both components of TWA 23 have SpT=M3.}

\end{deluxetable}

\begin{deluxetable}{lccccccccclll}
\tabletypesize{\scriptsize}
\rotate
\tablecaption{Kinematics of known TWA members observed with MIKE \label{table_twamembers_kinematics}}
\tablewidth{0pt}
\tablehead{
\colhead{Name} & \colhead{RA \& DEC}  &\colhead{pmRA\tablenotemark{a}}  &  \colhead{pmDec} &  \colhead{Dist.\tablenotemark{b}} & \colhead{RV} & \colhead{U} & \colhead{V} & \colhead{W} \\
\colhead{} & \colhead{deg.} & \colhead{mas~yr$^{-1}$} & \colhead{mas~yr$^{-1}$} &  \colhead{pc} & \colhead{km~s$^{-1}$} & \colhead{km~s$^{-1}$} & \colhead{km~s$^{-1}$} & \colhead{km~s$^{-1}$}

}
\startdata

TWA 2AB	&	167.31	-30.03	&	-95.5	$\pm$	2.9	&	-23.5	$\pm$	2.8	&	48	&	10.58	$\pm$	0.51	&	-14.9	$\pm$	1.7	&	-18.2	$\pm$	1.1	&	-7.7	$\pm$	1.4	\\
TWA 3Aab	&	167.62	-37.53	&	-100	$\pm$	7	&	-14	$\pm$	11	&	31	&	9.52	$\pm$	0.86	&	-10.0	$\pm$	1.7	&	-14.0	$\pm$	1.1	&	-3.9	$\pm$	1.6	\\
TWA 3B	&	167.62	-37.53	&	-100	$\pm$	7	&	-14	$\pm$	11	&	31	&	9.89	$\pm$	0.62	&	-9.9	$\pm$	1.6	&	-14.4	$\pm$	0.9	&	-3.8	$\pm$	1.6	\\
TWA 5Aab	&	172.98	-34.61	&	-85.3	$\pm$	3.6	&	-23.3	$\pm$	3.7	&	38	&	13.30	$\pm$	2.00	&	-8.5	$\pm$	1.4	&	-18.7	$\pm$	1.9	&	-2.5	$\pm$	1.3	\\
TWA 7	&	160.63	-33.67	&	-122.2	$\pm$	2.2	&	-29.3	$\pm$	2.2	&	34	&	12.21	$\pm$	0.24	&	-13.2	$\pm$	1.5	&	-17.7	$\pm$	0.7	&	-8.5	$\pm$	1.4	\\
TWA 8A	&	173.17	-26.87	&	-90	$\pm$	2	&	-20	$\pm$	15	&	44	&	8.34	$\pm$	0.48	&	-13.1	$\pm$	2.1	&	-15.8	$\pm$	1.6	&	-4.4	$\pm$	2.6	\\
TWA 8B	&	173.17	-26.87	&	-86	$\pm$	3	&	-22	$\pm$	38	&	27	&	8.93	$\pm$	0.27	&	-6.8	$\pm$	2.4	&	-12.7	$\pm$	2.1	&	-0.7	$\pm$	3.9	\\
TWA 10	&	188.77	-41.61	&	-78	$\pm$	23	&	-32	$\pm$	8	&	67	&	6.75	$\pm$	0.40	&	-15.7	$\pm$	6.6	&	-21.1	$\pm$	4.2	&	-8.6	$\pm$	2.7	\\
TWA 12	&	170.27	-38.75	&	-60	$\pm$	14	&	-12	$\pm$	25	&	63	&	13.12	$\pm$	1.59	&	-11.0	$\pm$	5.1	&	-19.1	$\pm$	2.8	&	-4.6	$\pm$	6.7	\\
TWA 14ab	&	168.36	-45.40	&	-43.3	$\pm$	2.6	&	-7	$\pm$	2.4	&	113	&	15.83	$\pm$	2.00	&	-14.4	$\pm$	2.3	&	-23.0	$\pm$	2.1	&	-8.1	$\pm$	1.8	\\
TWA 15B	&	188.59	-48.26	&	--			&	--			&	100	&	10.03	$\pm$	1.66	&	--			&	--			&	--			\\
TWA 16	&	188.73	-45.64	&	-53.2	$\pm$	5.2	&	-19	$\pm$	5.2	&	65	&	9.01	$\pm$	0.42	&	-8.5	$\pm$	1.9	&	-17.2	$\pm$	1.4	&	-4.0	$\pm$	1.7	\\
TWA 22AB	&	154.36	-53.91	&	-174.8	$\pm$	9	&	-13.6	$\pm$	9	&	20	&	13.57	$\pm$	0.26	&	-10.1	$\pm$	1.5	&	-16.3	$\pm$	0.4	&	-9.7	$\pm$	1.3	\\
TWA 23ab	&	181.86	-32.78	&	-44	$\pm$	7	&	-12	$\pm$	3	&	61	&	8.52	$\pm$	1.20	&	-7.0	$\pm$	2.1	&	-14.0	$\pm$	1.6	&	-1.0	$\pm$	1.1	\\

\enddata

\tablenotetext{a}{Proper motions are from the NOMAD catalog \citep{zach05}.}
\tablenotetext{b}{Photometric distances are calculated using the \cite{bara98} models with an age of 10 Myr, $T_{eff}$ from \cite{ment08} and taking into account binarity assuming equal flux components. Uncertainties are $\approx$ 10\%. Distances agree within error bars with the predicted values from \cite{mama05} and E.~E.~Mamajek (private communication), except TWA 15B which is predicted to be at 41 $\pm$ 6 pc, respectively.}																														

\end{deluxetable}

\begin{deluxetable}{lccccccccclll}
\tabletypesize{\scriptsize}
\rotate
\tablecaption{Spectral Age Diagnostics of TWA candidates observed with MIKE \label{table_spec}}
\tablewidth{0pt}
\tablehead{
\colhead{RA \& DEC} & \colhead{SpT}  &  \colhead{CaH-narrow} & \colhead{K I EW} & \colhead{Li EW} & \colhead{H$\alpha$ EW} & \colhead{H$\alpha$ 10\% width} &  \colhead{Youth} &  \colhead{Age\tablenotemark{b}} & \colhead{Note}\\
\colhead{$_{2MASS}$} & \colhead{M--} & \colhead{Index} & \colhead{\AA} & \colhead{\AA} & \colhead{\AA} & \colhead{km~s$^{-1}$} &  \colhead{Index\tablenotemark{a}} & \colhead{Myr}  & \colhead{}\\
\colhead{} & \colhead{($\pm 0.5$)}  &  \colhead{($\pm 0.03$)} & \colhead{($\pm 0.2$)} & \colhead{($\pm 0.05$)} & \colhead{($\pm 0.5$)} & \colhead{($\pm 4$)} & \colhead{} &  \colhead{} &  \colhead{} & \colhead{}

}
\startdata
							
10	13	21.43	-35	42	36.93	&	2.4	&	1.38	&	2.38	&	$<$0.1	&	-26.3	&	270	&	--	&	old	&	SB2, rapidly orbiting	\\
10	37	10.47	-35	05	01.52	&	3.2	&	1.40	&	1.49	&	$<$0.1	&	-6.1	&	108	&	1000	&	25 -- 300	&		\\
10	39	52.76	-35	34	03.03	&	0.6	&	1.18	&	0.82	&	$<$0.1	&	-1.5	&	87	&	1100	&	20 -- 110	&	VB (N)	\\
10	39	52.76	-35	34	03.03	&	0.3	&	1.16	&	0.83	&	$<$0.1	&	-1.5	&	90	&	1100	&	20 -- 110	&	VB (S)	\\
11	02	53.73	-31	45	10.57	&	2.3	&	1.38	&	1.67	&	$<$0.1	&	-3.2	&	79	&	0000	&	20 -- 300	&		\\
11	11	46.37	-39	37	34.78	&	2.2	&	1.52	&	1.09	&	$<$0.1	&	-12.4	&	124	&	1101	&	25 -- 130	&	SB2?\tablenotemark{c}	\\
11	11	52.67	-44	01	53.87	&	3.9	&	1.42	&	1.13	&	$<$0.1	&	-4.8	&	91	&	1100	&	90 -- 160	&		\\
11	18	12.37	-32	35	59.09	&	4.2	&	1.46	&	2.09	&	$<$0.1	&	-5.0	&	79	&	1000	&	90 -- 300	&		\\
11	30	53.55	-46	28	25.19	&	2.4	&	1.37	&	1.31	&	$<$0.1	&	-1.1	&	77	&	1100	&	20 -- 130	&		\\
11	31	14.83	-48	26	27.98	&	3.5	&	1.35	&	1.18	&	0.18	&	-7.3	&	233	&	1111	&	$\approx$15	&	LCC member	\\
12	03	08.07	-38	26	55.54	&	0.7	&	1.30	&	0.61, 0.26	&	$<$0.1	&	-1.9	&	95, 73	&	--	&	$>$300	&	SB2	\\
12	07	10.89	-32	30	53.72	&	4.2	&	1.35	&	1.35	&	0.41\tablenotemark{d}	&	-114.8	&	447	&	1111	&	10	&	TWA 31	\\
12	19	07.68	-41	01	57.81	&	2.9	&	1.35	&	1.46	&	$<$0.1	&	-3.3	&	72	&	1000	&	40 -- 300	&		\\
12	26	51.35	-33	16	12.47	&	6.3	&	1.39	&	1.21	&	0.60	&	-12.6	&	127	&	1111	&	10	&	VB, TWA 32	\\
12	33	43.57	-32	51	26.29	&	4.6	&	1.51	&	2.54	&	$<$0.1	&	-2.6	&	73	&	0000	&	180 -- 300	&		\\
12	45	43.60	-37	04	36.24	&	5.5	&	1.66	&	3.07	&	$<$0.1	&	-4.1	&	144	&	0000	&	180 -- 300	&		\\
\enddata

\tablenotetext{a}{Youth index in binary format in order of least to most restrictive age indicator: low-$g$ from CaH, low-$g$ from K i,  Li detection, accretion-level \ha\ emission.}																														
\tablenotetext{b}{~Lower limits on the stellar ages for early M dwarfs are provided for those stars with no lithium absorption ($\lambda$6708 \AA) using the lithium depletion time scales calculated by \cite{chab96}. However, it has been recently shown empirically, for at least the 12-Myr old $\beta$~Pic moving group, that lower ages limits of individual stars based on the the lack of lithium absorption systematically over-estimates the star's age  as compared to model isochrones \citep{yee10}.  This would imply that the lower age limits may be even lower, perhaps even as low as the accretion limit of $\approx$10 Myr. The upper age limit is set by the UV emission and/or low gravity using the evolution models of \cite{bara98}.}
\tablenotetext{c}{This star may be an unresolved SB2. See Section~\ref{kinematics} for more details.}
\tablenotetext{d}{The lower value of Li EW compared to other TWA members is likely due to veiling due to accretion (\citealt{dunc91}).}																														

\end{deluxetable}

\begin{deluxetable}{lcccccccccclll}
\tabletypesize{\scriptsize}
\rotate
\tablecaption{Kinematics of TWA candidates observed with MIKE \label{table_kinematics}}
\tablewidth{0pt}
\tablehead{
\colhead{RA \& DEC} & \colhead{pmRA}  &  \colhead{pmDec} & \colhead{REF\tablenotemark{a}} & \colhead{Dist.\tablenotemark{b}} & \colhead{RV} & \colhead{U} & \colhead{V} & \colhead{W} & \colhead{Note}\\
\colhead{$_{2MASS}$} & \colhead{mas~yr$^{-1}$} & \colhead{mas~yr$^{-1}$} & \colhead{} & \colhead{pc} & \colhead{km~s$^{-1}$} & \colhead{km~s$^{-1}$} & \colhead{km~s$^{-1}$} & \colhead{km~s$^{-1}$} & \colhead{}

}
\startdata
10	13	21.43	-35	42	36.93	&	18	$\pm$	6	&	-14	$\pm$	11	&	1	&	297	&	--			&	--			&	--			&	--			&	SB2	\\
10	37	10.47	-35	05	01.52	&	-52.8	$\pm$	2.4	&	8.9	$\pm$	2.4	&	2	&	41	&	12.45	$\pm$	0.54	&	-8.7	$\pm$	1.9	&	-13.7	$\pm$	0.7	&	0.6	$\pm$	0.7	&		\\
10	39	52.76	-35	34	03.03	&	-58.3	$\pm$	1.3	&	5.1	$\pm$	4.2	&	2	&	82	&	12.45	$\pm$	0.47	&	-19.1	$\pm$	4.0	&	-16.8	$\pm$	1.2	&	-5.1	$\pm$	2.3	&	VB (N)	\\
10	39	52.76	-35	34	03.03	&	-58.3	$\pm$	1.3	&	5.1	$\pm$	4.2	&	2	&	82	&	12.82	$\pm$	0.21	&	-19.1	$\pm$	4.0	&	-17.1	$\pm$	1.1	&	-5.0	$\pm$	2.3	&	VB (S)	\\
11	02	53.73	-31	45	10.57	&	22	$\pm$	3	&	4	$\pm$	4	&	1	&	218	&	95.27	$\pm$	0.22	&	28.6	$\pm$	4.9	&	-76.8	$\pm$	2.4	&	53.9	$\pm$	4.4	&		\\
11	11	46.37	-39	37	34.78	&	-34	$\pm$	12	&	2	$\pm$	12	&	1	&	200	&	-72.00	$\pm$	1.24	&	-43.8	$\pm$	11.2	&	55.9	$\pm$	4.6	&	-34.3	$\pm$	10.7	&	SB2?	\\
11	11	52.67	-44	01	53.87	&	-22	$\pm$	2	&	-12	$\pm$	4	&	1	&	34	&	17.64	$\pm$	0.30	&	2.1	$\pm$	0.6	&	-17.9	$\pm$	0.4	&	1.6	$\pm$	0.9	&		\\
11	18	12.37	-32	35	59.09	&	-38	$\pm$	3	&	8	$\pm$	8	&	1	&	106	&	-0.70	$\pm$	0.66	&	-18.6	$\pm$	4.3	&	-4.7	$\pm$	1.9	&	-3.7	$\pm$	3.4	&		\\
11	30	53.55	-46	28	25.19	&	-33.7	$\pm$	3.2	&	1.1	$\pm$	1.8	&	2	&	62	&	10.03	$\pm$	0.12	&	-5.6	$\pm$	1.9	&	-12.9	$\pm$	0.8	&	-0.3	$\pm$	0.8	&		\\
11	31	14.83	-48	26	27.98	&	-40.2	$\pm$	3.3	&	-5.8	$\pm$	1.5	&	2	&	93	&	17.02	$\pm$	1.16	&	-9.0	$\pm$	3.3	&	-22.6	$\pm$	1.9	&	-4.1	$\pm$	1.8	&	 LCC 	\\
12	03	08.07	-38	26	55.54	&	106.6	$\pm$	1.8	&	-18.3	$\pm$	1.1	&	2	&	45	&	-49.64	$\pm$	1.00	&	3.6	$\pm$	4.3	&	51.2	$\pm$	2.0	&	-19.0	$\pm$	0.5	&	SB2	\\
12	07	10.89	-32	30	53.72	&	-42	$\pm$	6	&	-36	$\pm$	3	&	1	&	110	&	10.47	$\pm$	0.41	&	-8.8	$\pm$	3.2	&	-25.5	$\pm$	2.5	&	-14.6	$\pm$	3.4	&	TWA 31	\\
12	19	07.68	-41	01	57.81	&	-74	$\pm$	20	&	34	$\pm$	4	&	1	&	101	&	6.44	$\pm$	0.11	&	-32.6	$\pm$	10.9	&	-18.2	$\pm$	5.3	&	12.9	$\pm$	3.0	&		\\
12	26	51.35	-33	16	12.47	&	-62.2	$\pm$	3.5	&	-24.7	$\pm$	3.9	&	2	&	53	&	7.15	$\pm$	0.26	&	-8.6	$\pm$	1.4	&	-15.7	$\pm$	1.1	&	-3.4	$\pm$	1.1	&	VB, TWA 32	\\
12	33	43.57	-32	51	26.29	&	-18.4	$\pm$	5.8	&	31.5	$\pm$	5.8	&	3	&	34	&	10.38	$\pm$	0.29	&	0.2	$\pm$	1.2	&	-7.4	$\pm$	0.7	&	9.4	$\pm$	1.2	&		\\
12	45	43.60	-37	04	36.24	&	21.7	$\pm$	9.6	&	-80.7	$\pm$	9.8	&	3	&	73	&	-10.81	$\pm$	1.82	&	8.2	$\pm$	4.1	&	2.3	$\pm$	2.8	&	-29.7	$\pm$	6.0	&		\\

\enddata

\tablenotetext{a}{Proper motion references: 1 = \citealt{zach05}, 2 = \citealt{zach10}, 3 = \citealt{roes10}}																														
\tablenotetext{b}{Photometric distances take into account youth (using \citealt{bara98}) and binarity assuming equal flux components. Uncertainties are $\approx$ 20\% for the non-TWA members and $\approx$10\% for the TWA and LCC members. See text for more details.}

\end{deluxetable}
 \clearpage


\end{document}